\documentclass[aps,prd,notitlepage,showpacs,nofootinbib,superscriptaddress]{revtex4-1}

\usepackage{graphicx}
\usepackage[utf8]{inputenc} 
\usepackage{amsmath}
\usepackage{amssymb}
\usepackage{float}
\usepackage{comment}
\usepackage{slashed}
\usepackage[normalem]{ulem}
\usepackage{array}
\usepackage{booktabs}

   \makeatletter
     \renewcommand\@make@capt@title[2]{%
      \@ifx@empty\float@link{\@firstofone}{\expandafter\href\expandafter{\float@link}}%
       {\textbf{#1}}\@caption@fignum@sep#2\quad}%
     \makeatother

\makeatletter 
\renewcommand{\fnum@figure}{\textbf{Fig.~\thefigure}}
\makeatother
\usepackage{multirow}
\usepackage{rotating}

\usepackage[usenames,dvipsnames]{color}
\usepackage[colorinlistoftodos]{todonotes}
\usepackage[colorlinks=true,citecolor=darkred,urlcolor=darkred, pdfborder={0 0 0}]{hyperref}
\usepackage[normalem]{ulem}

\definecolor{darkred}{rgb}{0.6,0,0}
\usepackage[colorinlistoftodos]{todonotes}

\definecolor{linkcolor}{rgb}{0,0,0.5}

\newcommand {\ignore}[1]{}

\def\gsim{\raise0.3ex\hbox{$\;>$\kern-0.75em\raise-1.1ex\hbox{$\sim\;$}}}
\def\lsim{\raise0.3ex\hbox{$\;<$\kern-0.75em\raise-1.1ex\hbox{$\sim\;$}}}

\providecommand{\be}{ \begin{equation} } 
\providecommand{\ee}{ \end{equation} }
\providecommand{\bea}{\begin{eqnarray}}
\providecommand{\eea}{\end{eqnarray}}

\providecommand{\to}{\rightarrow}

\usepackage{soul}

\definecolor{mightnightblue}{RGB}{25,25,112}

\definecolor{brown}{rgb}{0.59, 0.29, 0.0}

\def\vev#1{\left\langle #1\right\rangle}

\def\21{$\mathrm{SU(2)_L \otimes U(1)_Y}$}

\def\O{\hbox{$\cal O$ }}

\def\3311{$\mathrm{SU(3) \otimes SU(3)_L \otimes U(1)_X \otimes U(1)_{N}}$ }

\newcommand{\AddrAHEP}{%
  AHEP Group, Institut de F\'{i}sica Corpuscular --
  CSIC/Universitat de Val\`{e}ncia, Parc Cient\'ific de Paterna.\\
 C/ Catedr\'atico Jos\'e Beltr\'an, 2 E-46980 Paterna (Valencia) - Spain}
\newcommand{\Addraaa}{Ramakrishna Mission Residential College (Autonomous), Vivekananda Centre for Research, \\ Narendrapur, Kolkata, India}

\begin{document}

\title{\boldmath \color{BrickRed} 
Dynamical scoto-seesaw mechanism with gauged $B-L$ \\
}

\author{ Julio Leite }\email{julio.leite@ific.uv.es}
\affiliation{\AddrAHEP}

\author{Soumya Sadhukhan}\email{soumya.sadhukhan@rkmrc.in}
\affiliation{\Addraaa}
\affiliation{\AddrAHEP}

\author{Jos\'{e} W. F. Valle}\email{valle@ific.uv.es}
\affiliation{\AddrAHEP}

\begin{abstract}
\vspace{0.5cm}

We propose a dynamical scoto-seesaw mechanism using a gauged $B-L$ symmetry.
Dark matter is reconciled with neutrino mass generation, in such a way that the atmospheric scale arises \textit{a la seesaw},
while the solar scale is \textit{scotogenic}, arising radiatively from the exchange of ``dark'' states. This way we ``explain'' the solar-to-atmospheric scale ratio. 
The TeV-scale seesaw mediator and the two dark fermions carry different $B-L$ charges.
Dark matter stability follows from the residual matter parity that survives $B-L$ breaking. 
Besides having collider tests, the model implies sizeable charged lepton flavour violating (cLFV) phenomena, including Goldstone boson emission processes.

\end{abstract}

\maketitle
\noindent

\section{Introduction}

Despite the great success of the Standard Model (SM)~\cite{ParticleDataGroup:2022pth},
new physics is required in order to account for the existence of neutrino masses~\cite{McDonald:2016ixn,Kajita:2016cak} as well as dark matter~\cite{Bertone:2004pz}.
A popular paradigm for neutrino mass generation is the seesaw mechanism,
while weakly interacting massive particle (WIMP) dark matter candidates constitute a paradigm for explaining cold dark matter.
Even taking these paradigms for granted, there are many ways to realize either.
A particularly interesting possibility is provided by the so-called scotogenic approach~\cite{Ma:2006km,Hirsch:2013ola} in which WIMP dark matter mediates neutrino mass generation.
In the simplest schemes, all neutrino masses arise at the one-loop level, with a common overall scale, modulated only by Yukawa couplings.

Rather than invoking these paradigms separately, here we suggest a dynamical mechanism to realize naturally the scoto-seesaw scenario~\cite{Rojas:2018wym,Mandal:2021yph}
that reconciles dark matter (DM) and neutrino mass generation together\footnote{There are also attempts to realize the scoto-seesaw picture by combining it with family
  symmetries~\cite{Barreiros:2020gxu,Barreiros:2022aqu,Ganguly:2022qxj,Bonilla:2023pna}.}, opening the possibility of having a loop-suppressed solar-to-atmospheric scale ratio.

We first note that the $U(1)_{B-L}$ symmetry arises automatically in the SM and is closely related to neutrino masses.
However, the presence of non-vanishing anomaly coefficients forbids us to promote $U(1)_{B-L}$ to a local gauge symmetry.
Adding three lepton singlets $\nu_{iR}\sim -1$ cancels the anomaly coefficients, allowing for a local realisation.
Dirac neutrino mass entries arise in such a way that a $\nu_{iR}$-mediated (type-I) seesaw mechanism can be triggered by allowing for Majorana mass terms, which break $B-L$ by two units.
In such ``canonical'' construction the seesaw-mediating $\nu_{iR}$ carry identical charges so that all neutrino masses become proportional to a single energy scale.
As a result, this fails to account for the observed hierarchy $\Delta m^2_{sol}/\Delta m^2_{atm}$~\cite{deSalas:2020pgw}.

However, an alternative anomaly-free $U(1)_{B-L}$ can be obtained if instead we introduce three neutral fermions with $B-L$ charges $(f_{1R},f_{2R},N_{R})\sim(-4,-4,5)$~\cite{Montero:2007cd,Ma:2014qra}.

We show next that, thanks to their unequal charges, $f_{aR}$ and $N_{R}$ couple differently to the active neutrinos and may trigger different mass generation mechanisms,
providing a natural dynamical setup for the scoto-seesaw mechanism and an explanation for the smallness of the ratio $\Delta m^2_{sol}/\Delta m^2_{atm}$.
The neutrino-mass mediators $f_{aR}$ become part of a dark sector whose stability originates from an unbroken matter parity that survives the breaking of the gauged $U(1)_{B-L}$ symmetry. 

Besides having interesting dark matter features and collider phenomenology, an interesting implication of our model concerns charged lepton flavour violation.
Indeed, a characteristic feature to notice is the presence of sizeable cLFV processes involving Goldstone boson emission~\cite{Romao:1991tp,Hirsch:2009ee,Xing:2022rob},
in addition to conventional cLFV processes, such as those involving photon emission, for instance $\mu \to e \gamma$.

\section{Gauged $B-L$ extension of the Standard Model}
\label{sec:model}

Here, we describe the essential features of our proposal. 
The leptons and scalars present in the model and their symmetry properties are shown in Table \ref{Tab1}. 
In addition to the SM leptons, we introduce $f_{aR}$ ($a=1,2$) and $N_R$ -- singlets under the SM group but charged under $U(1)_{B-L}$. 
Besides a SM-like Higgs doublet $H$, the scalar sector contains two extra $SU(2)_L$ doublets ($\Phi$ and $\eta$) and four singlets ($\varphi_{1,2,3}, \sigma$). 
Note that all of the new fields are charged under $U(1)_{B-L}$.
The specific roles of the extra fields will become clear in the next sections when we discuss neutrino mass generation, dark matter, and how these are linked to each other in our model.

\begin{table}[h!]
\begin{center}
\begin{tabular}{| c || c | c | c |}
  \hline
&  \,\, Fields \,\, &  \,\, $SU(2)_L \otimes U(1)_Y$ \,\,  & \,\, $U(1)_{B-L}$ $\to$   $M_P$ \,\, \\
\hline \hline
\multirow{4}{*}{ \begin{turn}{90}\,\,\,\,\,\,Leptons \end{turn} } &
 $L_{iL}$     &    ($\mathbf{2}, {-1/2}$)       &   ${-1}$ $\to$  $+1$  \\
 & $e_{iR}$     &   ($\mathbf{1}, {-1}$)      & $-1$  $\to$  $+1$ \\
&   $N_{R}$       &   ($\mathbf{1}, {0}$)      & $5$   $\to$  $+1$ \\
\hline
\multirow{4}{*}{\begin{turn}{90} \,\,\,\,Scalars \,\,\,\,\,\,\,\,\,\,\,\,\,\, \end{turn} } &
  $H$  		 &  ($\mathbf{2}, {1/2}$)      &  ${0}$      $\to$  $+1$   \\
  &$\Phi$  		 &  ($\mathbf{2}, {1/2}$)      &  ${6}$   $\to$  $+1$     \\
  & $\varphi_1$             &  ($\mathbf{1}, {0}$)        &  ${10}$ $\to$  $+1$    \\
& $\varphi_2$             &  ($\mathbf{1}, {0}$)        &  ${8}$ $\to$ $+1$ \\
& $\varphi_3$             &  ($\mathbf{1}, {0}$)        &  ${2}$ $\to$  $+1$ \\
\hline \hline
\multirow{4}{*}{\begin{turn}{90}\,\,\, Dark\end{turn} } &
$f_{aR}$       &   ($\mathbf{1}, {0}$)      & $-4$ $\to$  $-1$  \\
\cmidrule(rl){2-4}
  &$\eta$  		 &  ($\mathbf{2}, {1/2}$)      &  ${-3}$  $\to$  $-1$      \\
  & $\sigma$          	 &  ($\mathbf{1}, {0}$)        &  ${9}$  $\to$  $-1$    \\
    \hline
  \end{tabular}
\end{center}
\caption{
  Lepton, scalar and dark sector fields with their symmetry transformation properties ($i=1,2,3$ and $a=1,2$). }
 \label{Tab1}
\end{table}

First we note that, with this field content, we ensure that $B-L$ is free from anomalies and hence can be promoted to local gauge symmetry. 
To see this, first recall that in the SM, the conventional $B-L$ appears as an automatic global symmetry that is anomalous due to the non-vanishing of the following triangle anomaly coefficients: 
\begin{equation} 
A_1^{\mbox{SM}}[U(1)_{B-L}^3] = -3 \quad\mbox{and}\quad A_2^{\mbox{SM}} [\mbox{Grav}^2 U(1)_{B-L}] = - 3.
\end{equation}
By extending the SM with the fermions $f_{aR}$ and $N_R$, as given in Table \ref{Tab1}, these coefficients vanish exactly \cite{Montero:2007cd,Ma:2014qra} 
\bea 
A_1[U(1)_{B-L}^3] &=& A_1^{\mbox{SM}}[U(1)_{B-L}^3] +A_1^{\mbox{new}}[U(1)_{B-L}^3] = - 3 - [(-4)^3+(-4)^3+(5)^3] = 0, \\
A_2 [\mbox{Grav}^2 U(1)_{B-L}] &=& A_2^{\mbox{SM}} [\mbox{Grav}^2 U(1)_{B-L}] +A_2^{\mbox{new}} [\mbox{Grav}^2 U(1)_{B-L}] = -3 - (-4-4+5) = 0.\nonumber
\eea
Needless to say, since these new fermions are singlets under the SM gauge group, they do not contribute to the other anomaly coefficients, {\it i.e.},
$U(1)_Y^2 U(1)_{B-L}$, $U(1)_Y U(1)_{B-L}^2$, $SU(2)_L^2 U(1)_{B-L}$ and $SU(3)_C^2 U(1)_{B-L}$ which, therefore, remain zero. 

With the fields in Table \ref{Tab1}, we can write down the most general renormalisable Yukawa Lagrangian
as\footnote{
  For convenience, we have omitted the Yukawa interactions involving quarks and $H$, which are standard.} 
\bea\label{eq:Yuk}
-\mathcal{L}_Y &=& Y^H_{ij} \overline{L}_{iL}H e_{jR} +Y^{\Phi} _{i} \overline{L}_{iL}\tilde{\Phi}N_{R}+ Y^{\eta} _{ia} \overline{L}_{iL}\tilde{\eta}f_{aR}+\frac{Y^{N}}{2}\varphi_1^* \overline{(N_{R})^c}N_{R}+\frac{Y^{f}_{a}}{2} \varphi_2 \overline{(f_{aR})^c}f_{aR}+h.c.,
\eea

Analogous to the SM case, once the Higgs doublet $H$ acquires its vacuum expectation value (VEV), the charged fermions become massive. 
In contrast, neutrino mass generation will involve the other scalar bosons, as discussed in Sec. \ref{sec:numass}.

In the scalar sector, we assume that only neutral fields with even $B-L$ charges acquire VEVs.
Thus, a subgroup of $U(1)_{B-L}$ remains conserved after spontaneous symmetry breaking takes place. The residual symmetry is matter parity and can be defined as  
\be\label{eq:MP}
M_P = (-1)^{3(B-L)+2s}.
\end{equation}
Therefore, only $f_{aR}$, $\eta$ and $\sigma$ are odd under $M_P$.
Due to $M_P$ conservation, the lightest of such fields is stable and, if electrically neutral, will be our dark matter candidate.

\section{Scalar sector} 
\label{sec:Sca}

For convenience, the most general renormalisable scalar potential is separated in two parts $V=V_1+V_2$, where 
\begin{eqnarray} 
 V_1 &=&  \sum_{i=1}^3 \left[ \mu_{\mathcal{D}_i}^2 \mathcal{D}_i^\dagger \mathcal{D}_i + \lambda_{\mathcal{D}_i} (\mathcal{D}_i^\dagger \mathcal{D}_i)^ 2 \right] + \sum^{i < j }_{i,j}\left[ \lambda_{\mathcal{D}_i \mathcal{D}_j} (\mathcal{D}_i^\dagger \mathcal{D}_i)(\mathcal{D}_j^\dagger \mathcal{D}_j) + \lambda^\prime_{\mathcal{D}_i \mathcal{D}_j} (\mathcal{D}_i^\dagger \mathcal{D}_j)(\mathcal{D}_j^\dagger \mathcal{D}_i) \right] \\
 && + \sum_{k=1}^4 \left[\mu_{\mathcal{S}_k}^2 S_k^\dagger \mathcal{S}_k + \lambda_{\mathcal{S}_k} (\mathcal{S}_k^\dagger \mathcal{S}_k)^2\right] +  \sum^{k < l }_{k,l} \lambda_{\mathcal{S}_k \mathcal{S}_l} (\mathcal{S}_k^\dagger \mathcal{S}_k)(\mathcal{S}_l^\dagger \mathcal{S}_l) + \sum_{i,k} \lambda_{\mathcal{D}_i \mathcal{S}_k}(\mathcal{D}_i^\dagger \mathcal{D}_i)(\mathcal{S}_k^\dagger \mathcal{S}_k),\label{eq:V1}\nonumber \\
-V_2 &=& \frac{\mu_1}{\sqrt{2}}\, \Phi^\dagger \eta \sigma + \frac{\mu_2}{\sqrt{2}} \varphi_1^* \varphi_2 \varphi_3  + \lambda_1\, \varphi_1\varphi_2 \sigma^{*\,2}  + \lambda_2 \Phi^\dagger H \varphi_2\varphi_3^*+ \text{h.c.}\,.\label{eq:V2}
\end{eqnarray}

where $\mathcal{D}_i=H,\Phi,\eta$ ($i=1,2,3$) represent the $SU(2)_L$ doublets and $\mathcal{S}_k=\varphi_1,\varphi_2,\varphi_3,\sigma$ ($k=1,2,3,4$) the singlets,
while $\mu_{H}^2$ and $\mu_{\varphi_i}^2$ are negative. The first part, $V_1$, contains only self-adjoint operators, while the second, $V_2$, includes all non-self-adjoint operators allowed by the symmetries. 

The scalars are decomposed as follows
\bea
H &=&\begin{pmatrix} H^+ \\ \frac{1}{\sqrt{2}}(v_H + S_H + i A_H) \end{pmatrix},  \quad \Phi =\begin{pmatrix} \Phi^+ \\ \frac{1}{\sqrt{2}}(v_\Phi + S_\Phi + i A_\Phi) \end{pmatrix},  \quad \eta =\begin{pmatrix} \eta^+ \\ \frac{1}{\sqrt{2}}(S_\eta + i A_\eta) \end{pmatrix},\\
\sigma &=& \frac{1}{\sqrt{2}}(S_{\sigma} + i A_{\sigma}), \quad
\varphi_i =\frac{1}{\sqrt{2}}(v_{\varphi_i} + S_{\varphi_i} + i A_{\varphi_i}), \quad \text{with} \,\,i=1,2,3,
\nonumber
\eea

and we assume that the breaking of $U(1)_{B-L}$ takes place well above the electro-weak scale: $v_{\varphi_i}^2\gg v_{EW}^2 = v_H^2 + v_\Phi^2$.
Notice that only the electrically neutral scalars with  even $U(1)_{B-L}$ charges acquire VEVs so that matter parity ($M_P$), defined in Eq. (\ref{eq:MP}), remains conserved.

After replacing the field decompositions above into the potential, we obtain the following tadpole equations 
\bea
 v_H \left[2 \mu_H^2+2 \lambda_H v_H^2+\lambda_{ H\varphi_1} v_{ \varphi_1}^2+\lambda_{ H\varphi_2} v_{ \varphi_2}^2+\lambda_{ H\varphi_3} v_{ \varphi_3}^2+ v_\Phi^2 (\lambda_{H\Phi}+\lambda_{H\Phi}^\prime)\right]-\lambda_2 v_\Phi v_{ \varphi_2} v_{ \varphi_3}&=&0\, ,\\
v_\Phi \left[2 \mu_\Phi^2 +2 \lambda_{\Phi} v_\Phi^2+\lambda_{\Phi \varphi_1} v_{ \varphi_1}^2+\lambda_{\Phi \varphi_2} v_{ \varphi_2}^2+\lambda_{\Phi \varphi_3} v_{ \varphi_3}^2+v_H^2  (\lambda_{H\Phi}+\lambda_{H\Phi}^\prime)\right]-\lambda_2 v_H v_{ \varphi_2} v_{ \varphi_3}&=&0\, ,\nonumber\\
v_{\varphi_1} \left(2 \mu_{\varphi_1}^2+2 \lambda_{\varphi_1} v_{ \varphi_1}^2+\lambda_{\Phi \varphi_1} v_\Phi^2+\lambda_{\varphi_1 \varphi_2} v_{ \varphi_2}^2+\lambda_{\varphi_1 \varphi_3} v_{ \varphi_3}^2+\lambda_{ H\varphi_1} v_H^2 \right)-\mu_2 v_{ \varphi_2} v_{ \varphi_3}&=&0\, ,\nonumber\\
v_{ \varphi_2}\left(2 \mu_{\varphi_2}^2+2 \lambda_{\varphi_2} v_{ \varphi_2}^2+ \lambda_{ H\varphi_2} v_H^2 +\lambda_{\Phi \varphi_2} v_\Phi^2 +\lambda_{\varphi_1 \varphi_2} v_{ \varphi_1}^2 \right)-v_{ \varphi_3}\left(\lambda_2 v_H v_\Phi +\mu_2 v_{ \varphi_1}\right) &=&0\, ,\nonumber\\
v_{ \varphi_3}\left(2 \mu_{\varphi_3}^2 +2 \lambda_{\varphi_3} v_{ \varphi_3}^2 +\lambda_{ H\varphi_3} v_H^2 +\lambda_{\Phi \varphi_3} v_\Phi^2 +\lambda_{\varphi_1 \varphi_3} v_{ \varphi_1}^2 \right)-v_{ \varphi_2}\left(\lambda_2 v_H v_\Phi + \mu_2 v_{ \varphi_1}\right)&=&0\,.  \nonumber
\eea
In the limit where $\mu_\Phi$ is much larger than the other mass scales in the model, the second tadpole equation leads to the induced VEV
\be\label{eq:vevPhi} 
v_\Phi \simeq \frac{\lambda_2 v_H v_{\varphi_2}v_{\varphi_3}}{2\mu_\Phi^2} \equiv v_H \epsilon\ll v_H.
\end{equation}
Therefore, the induced VEV of the doublet $\Phi$ is suppressed as in the type-II seesaw mechanism~\cite{Mandal:2022ysp,Mandal:2022zmy},
and also models with a leptophilic Higgs doublet~\cite{Ma:2000cc,Grimus:2009mm},
a natural example of which emerges naturally within the linear seesaw mechanism~\cite{Batra:2022arl,Batra:2023ssq,Batra:2023mds}.

Due to matter parity conservation, which survives as a remnant symmetry, fields with different $M_P$ charges remain unmixed.
Therefore, we can separate the scalar spectrum into a sector with fields transforming trivially under $M_P$, the $M_P$-even sector, and another with those that do not, the $M_P$-odd sector.
Moreover, for simplicity, we assume that all the VEVs as well as the couplings in $V$ are real so that CP is conserved. 

Starting with the $M_P$-even sector, we have a $5\times 5$ squared-mass matrix for the CP-even scalars $(S_H, S_\Phi, S_{\varphi_1}, S_{\varphi_2}, S_{\varphi_3})$, given in Eqs. (\ref{eq:MS2}) and (\ref{eq:MS2d}) in Appendix \ref{app:scalar}.
The associated physical states $S_{1,2,3,4,5}$ all become massive, and one of them is identified as the $125$ GeV Higgs boson, $S_1  \equiv h$ discovered at CERN~\cite{ATLAS:2012yve,CMS:2012qbp}. 
The other scalars are expected to be heavier, three of them, $S_{2,3,4}$, with masses proportional to the $B-L$-breaking scale $v_{\varphi_i}$. 
The mass of the heaviest state, $S_5$, will be governed by the largest scale in the model, {\it i.e.}, $\mu_\Phi$. Thus, in the limit of interest, $h\simeq S_H$ and $S_5 \simeq S_\Phi$. 

Concerning the CP-odd fields $(A_H, A_\Phi, A_{\varphi_1}, A_{\varphi_2}, A_{\varphi_3})$, there is another $5 \times 5$ squared-mass matrix, given in Eq. (\ref{eq:MA2}) in Appendix \ref{app:scalar}.
Three of the mass eigenstates are massless, and two of them can be written as
\begin{eqnarray}
G_Z &=& v_{EW}^{-1}\left( v_H A_H + v_\Phi A_\Phi \right) \\
G_{Z^\prime} &=& v_{BL}^{-1} \left( 6 v_\Phi A_\Phi + 10 v_{\varphi_1} A_{\varphi_1}+ 8 v_{\varphi_2} A_{\varphi_2} + 2 v_{\varphi_3} A_{\varphi_3}\right),\nonumber\\
\mbox{with}\quad v_{BL} &=& \sqrt{6^2 v_\Phi^2+10^2 v_{\varphi_1}^2+8^2 v_{\varphi_2}^2+2^2 v_{\varphi_3}^2},
\end{eqnarray}
and are absorbed by the neutral gauge bosons $Z$ and $Z^\prime$.

On the other hand, two states, $A_1$ and $A_2$, become massive\footnote{The exact mass expressions are given in Eq. (\ref{eq:MA12}) in Appendix \ref{app:scalar}.} and,
in the limit $v_{\varphi} \equiv v_{\varphi_i}\gg v_H \gg v_\Phi$, their masses can be written as 
\be
m^2_{A_1} \simeq \frac{3 \mu_2 v_\varphi}{2} \quad \mbox{and} \quad m^2_{A_2} \simeq \mu_\Phi^2.
\end{equation}

Finally, there is a remaining massless field, a physical Nambu-Goldstone boson, $G$, analogous to the Majoron~\cite{Chikashige:1980ui,Schechter:1981cv,Gonzalez-Garcia:1988okv}.

In order to obtain its profile, we make use of the fact that it must be orthogonal to would-be Goldstone bosons $G_{Z,Z^\prime}$,
as well as to the massive $A_{1,2}$ fields; see Appendix \ref{app:scalar}, Eq. (\ref{eq:G}).
Within the same limit as above, we find that this physical Goldstone lies mainly along the $SU(2)_L$ singlet directions,
while its projections along the doublets are suppressed by VEV ratios 
\be\label{eq:goldstone}
G\simeq \frac{1}{\sqrt{14}}\left(5\frac{v_\Phi^2}{v_H v_\varphi} A_H - 5\frac{v_\Phi}{v_\varphi} A_\Phi + A_{\varphi_1} - 2 A_{\varphi_2} + 3 A_{\varphi_3}\right).
\end{equation} 
This massless boson unveils the existence of a spontaneously broken accidental symmetry, identified in Appendix \ref{app:AccidentalU1}.

As we will comment below in Sec.~\ref{sec:Goldcoup}, there are stringent limits on the flavour-conserving Goldstone boson couplings to charged leptons, mainly electrons.
Moreover, as discussed in Sec.~\ref{sec:clfv}, there are also bounds on the flavour-violating Goldstone boson couplings to charged leptons.\\[-.2cm]

Turning now to the charged fields, $(H^\pm, \Phi^\pm )$, we have
\bea\label{eq:chargedS}
M^2_{\pm}=\frac{1}{2}
\left(
\begin{array}{cc}
 \lambda_2 \frac{v_{ \varphi_2} v_{ \varphi_3} v_\Phi}{v_H}-\lambda_{H\Phi}^\prime v_\Phi^2 &  -\lambda_2 v_{ \varphi_2} v_{ \varphi_3} + \lambda_{H\Phi}^\prime v_H v_\Phi\\
 -\lambda_2 v_{ \varphi_2} v_{ \varphi_3} + \lambda_{H\Phi}^\prime v_H v_\Phi & \lambda_2\frac{v_{ \varphi_2} v_{ \varphi_3}  v_H}{v_\Phi} -\lambda_{H\Phi}^\prime v_H^2\\
\end{array}
\right).
\eea
After diagonalising the above matrix, we find that one eigenstate is massless (absorbed by the $W^\pm$ gauge boson), whereas the other, $\phi^\pm$, gets a mass
\bea\label{eq:phip}
\phi^\pm =\frac{1}{\sqrt{v_H^2+v_\Phi^2}}\left(-v_{\Phi} H^\pm + v_{H} \Phi^\pm\right) \quad\text{with}\quad
m_{\phi^\pm}^2=\frac{(\lambda_2 v_{ \varphi_2} v_{ \varphi_3}-\lambda_{H\Phi}^\prime v_H v_\Phi)\left(v_H^2+v_\Phi^2\right) }{2 v_H v_\Phi}\,.
\eea
Notice that the dominant component of $\phi^{\pm}$ lies along the doublet $\Phi$, which gets a large mass $m_{\phi^\pm}\simeq \mu_\Phi$; see Eq.~(\ref{eq:vevPhi}). \\[-.2cm]

Turning now to the $M_P$-odd sector, the CP-even and CP-odd scalars have the following mass matrices, respectively, 
\bea
M^{2}_{a(s)} &=&\frac{1}{2}\left(
\begin{array}{cc}
 X & \kappa_{a(s)}  \mu_1v_\Phi \\
 \kappa_{a(s)}  \mu_1v_\Phi  & Y_{a(s)}  \\
\end{array}
\right),\\
\text{with}  \quad  X &=& 2\mu_\eta^2 +(\lambda_{ H\eta}+\lambda_{ H\eta}^\prime) v_H^2+(\lambda_{\Phi\eta} +\lambda_{\Phi \eta}^\prime)v_\Phi^2 +  \lambda_{ \eta \varphi_1}v_{ \varphi_1}^2 + \lambda_{ \eta \varphi_2}v_{ \varphi_2}^2 +  \lambda_{ \eta \varphi_3}v_{ \varphi_3}^2,\nonumber\\
Y_{a(s)} &=& 2\mu_\sigma^2+\lambda_{H \sigma} v_H^2+ \lambda_{\Phi\sigma}v_\Phi^2  +\lambda_{\varphi_1 \sigma }v_{ \varphi_1}^2   + \lambda_{\varphi_2 \sigma }v_{ \varphi_2}^2  + \lambda_{\varphi_3 \sigma }v_{ \varphi_3}^2 +\kappa_{a(s)} 2\lambda_1 v_{ \varphi_1} v_{ \varphi_2} \,\nonumber\\
\mbox{and} \quad\kappa_{a(s)}&=& +1(-1)\nonumber .
\eea
when expressed in the bases $(S_\eta, S_\sigma)$ and $(A_\eta, A_\sigma)$. The matrices can be diagonalised by performing the following rotations
\bea
\label{eq:mix}
\begin{pmatrix} a(s)_1 \\ a(s)_2  \end{pmatrix} =  \begin{pmatrix} \cos{\theta_{a(s)}} & -\sin{\theta_{a(s)}} \\ \sin{\theta_{a(s)}} & \cos{\theta_{a(s)}}  \end{pmatrix} \begin{pmatrix} A(S)_\eta \\ A(S)_\sigma  \end{pmatrix},\quad\mbox{where}\quad\tan(2 \theta_{a(s)}) = \frac{\kappa_{a(s)} 2 \mu_1 v_\Phi}{Y_{a,s}-X}.
\eea
The corresponding eigenvalues are then given by  
\bea
m^{2}_{a_i(s_i)} = \frac{X+ Y_{a(s)}  \pm \mbox{sign}(X-Y_{a(s)})\sqrt{ (X-Y_{a(s)})^2+16 \mu_1^2 v_\Phi^2}}{4}.
\eea

Finally, the only charged scalar in the dark sector gets the following mass 
\bea\label{eq:metapm}
m_{\eta^\pm}^2=
\mu_\eta^2 +\frac{1}{2}\left(\lambda_{ H\eta} v_H^2+\lambda_{\Phi\eta}  v_\Phi^2+\lambda_{ \eta \varphi_1} v_{ \varphi_1}^2+\lambda_{ \eta \varphi_2} v_{ \varphi_2}^2+\lambda_{ \eta \varphi_3} v_{ \varphi_3}^2\right)\,.
\eea

\section{Gauge sector: $Z-Z^\prime$ mixing} 
\label{sec:gauge}

As usual, to find the gauge sector spectrum, we expand the covariant derivative terms in the scalar sector, {\it i.e.}, $\mathcal{L}\supset (D_\mu \phi_i)^\dagger(D^\mu \phi^i)$,
where $\phi_i$ denote the scalars in the model and\footnote{Given that the model contains ``large'' $B-L$ charges, $q_{BL}$, to ensure perturbativity of the associated gauge
  interactions, one may adopt the conservative limit $q_{BL}^{max}\times g^\prime = 10\times g^\prime \lesssim 1$, leading to $g^\prime \lesssim 0.1$.} 
\be
D_\mu \phi_i = \left(\partial_\mu - i g_L \frac{\sigma_a}{2} W^a_\mu - i g_Y Y B^Y_\mu - i g^\prime q_{BL} B^\prime_\mu\right)\phi_i.
\end{equation}

 When the scalars acquire VEVs, the neutral gauge bosons acquire the following squared-mass matrix, written in the basis
 $(W^3_\mu, B^Y_\mu, B^\prime_\mu),$\footnote{We are assuming, for simplicity, that the kinetic mixing between the Abelian fields vanishes.}
 \be
M^2=\frac{1}{4}\left(
\begin{array}{ccc}
 g_L^2 v_{EW}^2 & -g_L g_Y v_{EW}^2 & -12 g_L g^\prime v_\Phi^2 \\
 -g_L g_Y v_{EW}^2 & g_Y^2 v_{EW}^2 & 12 g^\prime g_Y v_\Phi^2 \\
 -12 g_L g^\prime v_\Phi^2 & 12 g^\prime g_Y v_\Phi^2 & 4 g^{\prime 2} v_{\varphi_1}^2\alpha \\
\end{array}
\right),
\end{equation}
where $\alpha = (2/v_{\varphi_1})^{2}\left(9v_\Phi^2 + 25 v_{\varphi_1}^2 + 16 v_{\varphi_2}^2 + v_{\varphi_3}^2\right)$.
In order to determine the mixing among the fields, we diagonalise this matrix in two steps.
First, we single out the photon ($A_\mu$) by making use of the mixing matrix $R_1$, and then
we diagonalise the resulting $Z-Z^\prime$ matrix with the help of $R_2$, namely
\be
R =
R_2\times R_1=
\left(
\begin{array}{ccc}
 1 & 0 & 0 \\
 0 & \cos{\delta} & -\sin{\delta} \\
 0 & \sin{\delta} & \cos{\delta} 
\end{array}
\right)\left(
\begin{array}{ccc}
 \frac{g_Y}{g} & \frac{g_L}{g} & 0 \\
 -\frac{g_L}{g} & \frac{g_Y}{g} & 0 \\
 0 & 0 & 1 \\
\end{array}
\right),\quad \mbox{with} \quad \tan(2\delta) = \frac{12 g g^\prime v_\Phi^2}{4  g^{\prime 2} v_{\varphi_1}^2\alpha-g^2 v_{EW}^2},
\end{equation}
and $g^2 = g_L^2 + g_Y^2$. 
It is easy to see that the $Z-Z^\prime$ mixing, parametrised by $\delta$, is rather suppressed since $v_\Phi/v_{\varphi_1}\ll 1$.
Thus, the mass eigenstates are given by 
\bea
A_\mu &=& \frac{g_Y}{g} W^3_{\mu} + \frac{g_L}{g} B^Y_{\mu},\\ 
Z_\mu &=&\frac{\cos{\delta}}{g}\left(g_L W^3_{\mu} +g_Y B^Y_{\mu} \right)-\sin{\delta} B^{\prime}_\mu,\nonumber\\ 
Z^\prime_\mu &=&\frac{\sin{\delta}}{g}\left(g_L W_{3\mu} +g_Y B_{\mu} \right)+\cos{\delta} B^\prime_\mu,\nonumber 
\eea
and the non-vanishing eigenvalues are 
\be
m^2_{Z,Z^\prime } = \frac{1}{8} \left[4  g^{\prime 2} v_{\varphi_1}^2 \alpha+g^2 v_{EW}^2 \mp \sqrt{(4  g^{\prime 2} v_{\varphi_1}^2\alpha -g^2 v_{EW}^2)^2 + (24 g g^\prime v_\Phi^2)^2}\right],
\end{equation}
which, upon assuming $v_{\varphi} \equiv v_{\varphi_i}\gg v_H\gg v_\Phi$, give $m_Z \simeq g v_{EW}/2$, while $m_{Z^\prime} \simeq 13 g^\prime v_{\varphi}$ is the mass of the heavy ${Z^\prime}$. 

\section{Dynamical scoto-seesaw mechanism} 
\label{sec:numass}

\begin{figure}[h!]
\centering
\includegraphics[scale=.9]{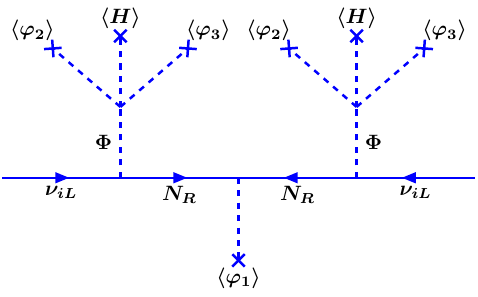}~~~~~
\includegraphics[scale=.9]{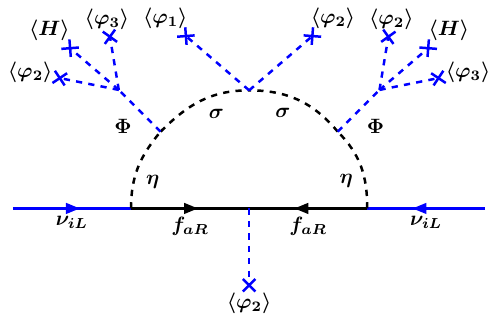}
\caption{
  Seesaw and scotogenic contributions to neutrino masses.
  The fields that are even under $M_P$ are depicted in blue, whereas the black legs represent the $M_P$-odd (dark) fields.}
\label{fig:massdiags}
\end{figure}

Neutrino masses arise from the operators in Eqs. (\ref{eq:Yuk}) and (\ref{eq:V2}) and are generated by the diagrams in Fig. \ref{fig:massdiags},
where the blue and the black legs denote the exchange of $M_P$-even and $M_P$-odd/dark fields, respectively. 

The tree-level contribution, in the basis $(\nu_{iL}, (N_R)^c)$, leads to the following mass matrix 
\bea
M^{\nu,N} = \frac{1}{\sqrt{2}}\begin{pmatrix} 
0 & 0 & 0 & Y^{\Phi} _{1}  v_\Phi \\
0 & 0 & 0 & Y^{\Phi} _{2}  v_\Phi\\
0 & 0 & 0 & Y^{\Phi} _{3}  v_\Phi\\
Y^{\Phi} _{1}  v_\Phi & Y^{\Phi} _{2}  v_\Phi & Y^{\Phi} _{3}  v_\Phi & Y^N v_{\varphi_1}\\                        
\end{pmatrix}.
\eea
Upon diagonalisation we find the seesaw-suppressed mass matrix for the active neutrinos to be
\be\label{eq:numass0}
M^{\nu(\mathrm{SS})}_{ij} \simeq -\frac{Y^{\Phi} _i Y^{\Phi} _j}{2} \frac{v_\Phi^2}{m_N},
\end{equation}
in the limit $v_{\varphi_i}\gg v_H \gg v_\Phi$.
Here $m_N \simeq v_{\varphi_1} Y^N/\sqrt{2}$ is the mass of $N_{R}$ and $ v_\Phi= v_H\epsilon $ is the small induced VEV defined in Eq. (\ref{eq:vevPhi}).  
The light neutrino mass matrix has only one non-vanishing eigenvalue $\sim -\frac{v_\Phi^2}{m_N}\sum_i (Y^{\Phi} _{i})^2$. 

One sees that, in contrast to the conventional type-I seesaw mechanism, in which neutrino mass suppression follows from the large size of $m_N$
with respect to the electroweak scale ($v_{EW}\sim v_H$),
here, there is an additional suppression from the small induced VEV of $\Phi$, characterized by $\epsilon$ given in Eq. (\ref{eq:vevPhi}).
This feature allows us to have moderate values for $m_N$, or $v_{\varphi_1}$, say around the TeV scale, without the need for appealing to tiny Yukawa couplings.  

At this point, we also stress that, in contrast to low-scale inverse~\cite{Mohapatra:1986bd,Gonzalez-Garcia:1988okv} or linear seesaw schemes~\cite{Akhmedov:1995ip,Akhmedov:1995vm,Malinsky:2005bi},
the presence of TeV-scale neutrino-mass-mediators does not require the addition of extra gauge singlets beyond the ``right-handed neutrinos''.
Moreover, since $m_N$ is directly associated with the spontaneous breaking of the gauged $B-L$, the mass of the associated gauge boson, $Z^\prime$,
can also be naturally within experimental reach, leading to rich phenomenological implications, as we discuss below.\\[-.2cm] 

Concerning the other two neutrinos, their masses are generated at one loop through the scotogenic mechanism, namely 
\begin{eqnarray}\label{scotomass}
M^{\nu(\mathrm{SC})}_{ij} &=& \sum_{c=1}^2 Y^{\eta} _{ic} \mathcal{M}_c\, Y^{\eta} _{jc},\quad\quad\mbox{with}\\
\mathcal{M}_{c} &=& \frac{m_{f_c}}{16 \pi^2} 
\left[ \frac{\cos^2 \theta_s m^2_{s_1}}{m^2_{s_1} - m_{f_c}^2} \ln \frac{m^2_{s_1}}{m_{f_c}^2} 
- \frac{\cos^2 \theta_a m^2_{a_1}}{m^2_{a_1} - m_{f_c}^2} 
\ln \frac{m^2_{a_1}}{m_{f_c}^2} + \frac{\sin^2 \theta_s m^2_{s_2}}{m^2_{s_2} - m_{f_c}^2} \ln \frac{m^2_{s_2}}{m_{f_c}^2} - \frac{\sin^2 \theta_a m^2_{a_2}}{m^2_{a_2} - m_{f_c}^2} 
\ln \frac{m^2_{a_2}}{m_{f_c}^2} \right],\nonumber
\end{eqnarray}
where $m_{f_c} = v_{\varphi_2} Y^f_c/\sqrt{2}$ are the masses of the two dark fermions $f_{cR}$.
One can easily check that when either $\lambda_1\to 0$ or $\mu_1\to 0$, the loop-generated masses vanish. 
When $\lambda_1\to 0$, $m_{s_i}\to m_{a_i}$, with $\cos^2\theta_s \to\cos^2\theta_a$ and $\sin^2\theta_s \to\sin^2\theta_a$,
leading to a cancellation between the first and the second, as well as the third and the fourth terms in Eq. (\ref{scotomass}).
On the other hand, when $\mu_1\to 0$, we have $\theta_s,\theta_a \to 0$ so that only the first and the second terms in Eq. (\ref{scotomass}) survive;
nevertheless, these cancel out since, in this limit, $m_{s_1}\to m_{a_1}$.

As usual in scotogenic setups, the loop mediators -- black fields in Fig. \ref{fig:massdiags} -- are part of a dark sector from which the lightest field is stable.
This can play the role of a WIMP dark matter candidate.  
In our case, the viable dark matter candidates are the neutral fermions $f_{cR}$ and the neutral scalars $s_i, a_i$.
The stability of the lightest of them is ensured by the residual subgroup of our gauged $B-L$ symmetry, {\it i.e.}, the conserved matter parity ($M_P$), defined in Eq. (\ref{eq:MP}). 

Lastly, since one neutrino mass is generated at tree level, while the other two masses arise through the scotogenic diagram at the one-loop level,
we expect these masses to be loop-suppressed with respect to the former~\cite{Rojas:2018wym,Mandal:2021yph}. 
Therefore, this framework favours the normal ordering of neutrino masses -- also preferred experimentally~\cite{deSalas:2020pgw} --
as well as an understanding of the origin of the smallness of the ratio $\Delta m^2_{sol}/\Delta m^2_{atm}$.

\section{Goldstone couplings to fermions}
\label{sec:Goldcoup}
The effective interaction between the Goldstone boson, $G$, and the fermions, $F_i$, can parametrised as 
\be\label{eq:Gff}
\mathcal{L}_{G F_i F_j} =  G\, \overline{F_{j}}\left(\Sigma_L^{GF\,ji}P_L+\Sigma_R^{GF\,ji}P_R\right)F_{i} + \text{h.c.},
\end{equation}
where $P_{L,R}$ are the usual chiral projectors, and $\Sigma^{GF}_{L,R}$ are model-dependent dimensionless coefficients. 
As seen from Eq. (\ref{eq:goldstone}), the Goldstone boson has projections along of the $M_P$-even scalars, including the SM-like Higgs doublet $H$.
As a result, $G$ couples to all of the fermions at tree level.

For quarks and charged leptons, which get their tree-level masses exclusively from their interactions with $H$, the tree-level couplings to $G$ arise from the projection of $G$ into the $H$, $\vev{ G|H}$.
This can be expressed as 
\bea\label{eq:SigmaGf}
|\Sigma_R^{GF\,ji}|= |\vev{ G | H }|\frac{ m_{F_i} }{v_H} \delta_{ij},
\eea
with $F_i=e,d,u$ representing the charged leptons, down- and up-type quarks, respectively. Notice that when we diagonalize the charged-fermion mass matrices, the G-couplings also become diagonal. 
In other words, flavour-violating couplings to charged fermions are absent at tree level. 

Of particular interest are the couplings to charged leptons, for example electrons.
Depending on the strength of such couplings, Goldstone bosons can be over-produced in Compton-like processes $e+\gamma \to e + G$ and emitted by stars leading to excessive stellar cooling.
 Recently it was noted that also the pseudoscalar couplings to the muon can be restricted by SN1987A cooling rates~\cite{Bollig:2020xdr}.
 The bounds for the couplings to electrons and muons are $|\Sigma^{G\,ee}|\lesssim 2.1\times 10^{-13}$ and $|\Sigma^{G\,\mu\mu}|\lesssim 2.1\times 10^{-10}$, respectively~\cite{Fontes:2019uld,Bollig:2020xdr}. 
 Assuming $v_{\varphi} = v_{\varphi_i}\gg v_H \gg v_\Phi$ Eq. (\ref{eq:goldstone}) leads to $v_\Phi^2/(v_H v_\varphi) < 7.5\times10^{-8}$ or, equivalently, 
\be
\frac{v_\Phi^2}{v_\varphi} \lesssim 1.9\times 10^{-5} \,\text{GeV}.
\end{equation}
One sees that both constraints can be satisfied through the above restriction on the Goldstone projection into the SM-like Higgs doublet.  
It is worth noticing that this effective mass scale, $v_\Phi^2/v_\varphi$ also appears in the seesaw-suppressed neutrino mass in Eq. (\ref{eq:numass0}), and should therefore be small. 

Note also that the Goldstone couplings to charged leptons also receive radiative contributions, mediated mainly by the dark sector, via the diagram in Fig. \ref{fig:cLFVdiagramsgG} (right). 
Such contributions can be calculated using the results of the next section by taking the diagonal entries of the one-loop couplings $\Sigma^G_{L,R}$ in Eqs. (\ref{eq:LmueX}) and (\ref{eq:mu2eG2}). 

The tree-level Goldstone couplings to neutrinos emerge from the second term in Eq. (\ref{eq:Yuk}) involving the leptophilic doublet $\Phi$, {\it i.e.}, $Y^\Phi_i \overline{L_{iL}} \tilde{\Phi} N_R$.
As a consequence, they are suppressed by the product of the Goldstone projection into $\Phi$, of order $v_\Phi/v_\varphi$, and the small $\nu-N$ mixing, also of order $v_\Phi/v_\varphi$.
 Dark-mediated loop contributions are also present; however, these are expected to be very small as they are proportional to neutrino masses, a feature
  reminiscent of Majoron models~\cite{Schechter:1981cv,Gonzalez-Garcia:1988okv}. 
  
\section{Charged lepton flavour violation}
\label{sec:clfv}

In this section, we discuss the most relevant charged lepton flavour violating (cLFV) processes occurring in our model, involving the first two families.
For theoretical and experimental reviews, see, for instance, Refs.~\cite{Cei:2014jtm,Lindner:2016bgg,Calibbi:2017uvl}.

An interesting characteristic feature to notice is the presence of charged lepton flavour violation involving Goldstone boson emission.
The projected $\mu \to e$ conversion at COMET \cite{COMET:2018auw} can also probe $\mu \to e G$.
There are good prospects for improving the sensitivities on Goldstone boson searches with the COMET experiment,
  where the sensitivity for the cLFV process $ \mu \to eG $ may improve from the current limit  ${BR}(\mu \to eG)=2.3\times 10^{-5} $ 
  in Phase-I to a sensitivity $ \O(10^{-8}) $ in Phase-II, under somewhat optimistic assumptions~\cite{Xing:2022rob}\footnote{{ In a similar fashion,
    the Mu3e experiment, which aims at substantially improving the bounds on the decay $\mu \to 3e$, can also probe $\mu \to e G$ \cite{Perrevoort:2018ttp}, with sensitivity
    one order of magnitude worse than the above one for COMET.}  }.
In what follows we pay special attention not only to radiative cLFV processes involving photon emission, but also to Goldstone boson emitting processes.

\subsection{cLFV through photon and Goldstone boson emission} 
\label{sec:clfv-through-photon}

\begin{figure}[h!]
\centering
\includegraphics[scale=0.9]{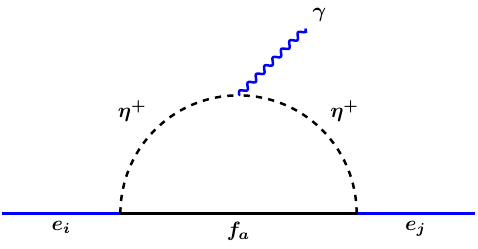}~~~~~
\includegraphics[scale=0.9]{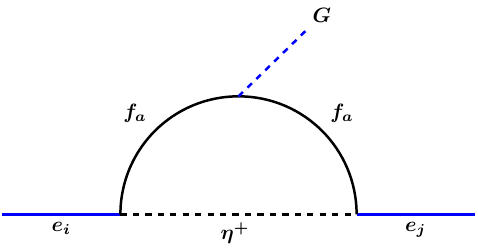}
\caption{
  Leading contributions to the charged lepton flavour violating decays  $e_i \to e_j \gamma$ and $e_i \to e_j G$. }
\label{fig:cLFVdiagramsgG}
\end{figure}

We focus on cLFV decays of the form $e_i \to e_j X$ -- with $X=\gamma, G$ -- which,
here, receive their main contributions from diagrams mediated by the $M_P$-odd fields in the ``dark'' sector, as shown in Fig. \ref{fig:cLFVdiagramsgG}. 

Other diagrams, although present, give rise only to sub-leading contributions. 
For instance, analogously to Fig. \ref{fig:cLFVdiagramsgG}, diagrams mediated by the $M_P$-even fields $\phi^+$ and $N$, instead of the $M_P$-odd fields $\eta^+$ and $f_a$, exist but their contributions are significantly suppressed by the large mass of $\phi^+$.
Likewise, charged-current contributions mediated by the $W^+$ gauge boson and $N$ are present; however, they are also suppressed due to the small $\nu-N$ mixing.

The effective operators associated with the decays $e_i \to e_j X$, with $X=\gamma,G$, can be expressed, respectively, as follows~\cite{Lavoura:2003xp,Escribano:2020wua}
\be\label{eq:LmueX}
\mathcal{L}_{X e_i e_j} = \frac{F_{\mu\nu}}{2}\overline{e_j}\sigma^{\mu\nu}\left(\Sigma_L^{\gamma\,ji} P_L+\Sigma_R^{\gamma\,ji} P_R\right) e_i  +  G\, \overline{e_{j}}\left(\Sigma_L^{G\,ji}P_L+\Sigma_R^{G\,ji}P_R\right)e_{i} + \text{h.c.},
\end{equation}
where $\sigma^{\mu\nu} = [\gamma^\mu,\gamma^\nu]$ and $P_{L,R}$ are the usual chiral projectors. 
Here $\Sigma^\gamma_{L,R}$ ($\Sigma^G_{L,R}$) are model-dependent coefficients with dimension $\mathrm{mass}^{-1}$ ($\mathrm{mass}^{0}$).

The branching ratios for the processes of interest, assuming $m_j/m_i\ll 1$, are given by
\cite{Lavoura:2003xp,Toma:2013zsa,Lindner:2016bgg,Babu:2007sm,Escribano:2020wua,Portillo-Sanchez:2023kbz}
\begin{eqnarray}\label{eq:mu2eg}
\mathrm{BR}\left(e_{i}\to e_{j}\gamma\right)\simeq \frac{m_{e_i}^3}{16 \pi \Gamma_{e_i}}\left(|\Sigma_L^{\gamma\,ji}|^2+|\Sigma_R^{\gamma\,ji}|^2\right)\quad \mbox{and} \quad
\mathrm{BR}\left(e_{i}\to e_{j}G\right)\simeq
\frac{m_{e_i}}{32 \pi \Gamma_{e_i}} \left(|\Sigma_L^{G\,ji}|^2+|\Sigma_R^{G\,ji}|^2 \right)\,.
\end{eqnarray}
In particular, we shall focus on the most constraining processes, {\it i.e.}, $\mu \to e \gamma$ and $\mu \to e G$,  
whose current limits are $\mathrm{BR}(\mu \to e \gamma) \lesssim 4.2 \times 10^{-13}$ \cite{MEG:2016leq} and $\mathrm{BR}(\mu \to e G) \lesssim 10^{-5}$~\cite{Jodidio:1986mz,Hirsch:2009ee,TWIST:2014ymv},
while future experiments are expected to improve these bounds to $\mathrm{BR}(\mu \to e \gamma) \lesssim 6\times 10^{-14}$ \cite{MEGII:2021fah}
and $\mathrm{BR}(\mu \to e G) \lesssim 10^{-8}$~\cite{Xing:2022rob}.

Taking into account that $m_e/m_\mu\ll 1$, the leading contribution to the first process can be approximated to 
\be\label{eq:mu2eg2}
\mathrm{BR}\left(\mu\to e\gamma\right) \simeq \frac{\alpha_\mathrm{em} m_\mu^5}{4\, m_{\eta^+}^4  \Gamma_{\mu}}\left|\left(Y^{\eta}F_1 Y^{\eta\dagger} \right)^{e\mu }\right|^2\, \quad
\text{with}\quad (F_1)_{ab} = \frac{1-6x_a+3x_a^2+2x_a^3-6x_a^2 \log x_a}{6(4\pi)^2(1-x_a)^4}\delta_{ab}\,,
\end{equation}
where $x_a=(m_{f_a}/m_{\eta^+})^2$, $\alpha_{\mathrm{em}}$ is the fine-structure constant, and $m_\mu \simeq 105.658$ MeV is the muon mass.
Here, $\Gamma_\mu = \Gamma_\mu^{SM} + \Gamma_\mu^{new} \simeq 3\times 10^{-19}\, \text{GeV} +\Gamma_\mu^{new}$ is the total decay width of the muon,
with $\Gamma_\mu^{new}$ representing the new contributions to the muon decay width calculated in this section.
Similarly, for the process involving the Goldstone boson emission, we have
\begin{equation} \label{eq:mu2eG2}
\mathrm{BR}\left(\mu\to e G\right)\simeq
\frac{ m_{\mu}^3 \, |\langle G|\varphi_2\rangle|^2 }{ 32 \pi\, v_{\varphi_2}^2 \Gamma_{\mu} } \left|\left(Y^{\eta}  F_2 Y^{\eta \dagger} \right)^{e\mu}\right|^2, \quad
 \mathrm{with }\quad(F_2)_{ab} =  \frac{x_a\left( x_a - 1 -\log x_a\right)}{2(2\pi)^2(x_a-1)^2}\delta_{ab},
\end{equation} 
where $\vev{ G|\varphi_2}$ is the Goldstone projection along $\varphi_2$. 

We would like to point out also that, despite the presence of several new neutral scalars as well as an extra neutral gauge boson ($Z^\prime$),
flavour-changing neutral currents involving charged leptons only take place at loop level.
Indeed, notice that, at tree level, two charged leptons can only couple to a scalar via the first term in Eq. (\ref{eq:Yuk}), governed by the $Y^H$ matrix.
The very same term leads to the charged lepton mass matrix: $M^e = Y^H \vev{ H }$. Therefore, by diagonalising $M^e$, the charged lepton couplings to neutral
scalars are also automatically diagonalised.
Lastly, since all the charged leptons couple to $Z^\prime$ with the same $B-L$ charge ($-1$), and there is no non-standard charged lepton,
the tree-level couplings between the charged leptons and $Z^\prime$ are also diagonal.

\subsection{Numerical results} 
\label{sec:numerical}

Having discussed the main cLFV processes, as well as the relevant bounds, we now proceed to calculate numerically the relevant contributions.
Note that the Yukawa couplings controlling the cLFV processes are also involved in the generation of neutrino masses.
Thus, in order to estimate the cLFV contributions, we need to choose as inputs benchmarks satisfying neutrino oscillation data.
Although we perform this task numerically, it is useful to develop an analytical form for extracting the Yukawa couplings from measured parameters that satisfy neutrino oscillation restrictions, {\it \`a la} Casas-Ibarra \cite{Casas:2001sr},
as this will optimise our subsequent scanning procedure.
For this purpose, we write the sum of the tree-level (seesaw) and the one-loop (scotogenic) neutrino masses in Eqs. (\ref{eq:numass0}) and (\ref{scotomass}) in the following compact form
\bea\label{eq:redef}
M^{\nu} &=& M^{\nu(\mathrm{SS})}+M^{\nu(\mathrm{SC})} = -Y \mathcal{X_M} \,Y^T,\quad \text{with}
\quad Y_{3\times 3}= \begin{pmatrix} Y^{\eta} _{3\times2} & Y^{\Phi} _{3\times1}
\end{pmatrix} \quad \text{and} \quad \mathcal{X_M}_{ij}=0\,\, (i\neq j),\\
 \mathcal{X_M}_{cc} &=& -\mathcal{M}_c, \quad \text{for}\quad c=1,2, \quad\mbox{and} \quad
\mathcal{X_M}_{33} = \frac{v_\Phi^2 }{2 m_N},\nonumber 
\eea
with $\mathcal{M}_c$ defined in Eq. (\ref{scotomass}). This way the ansatz below can be used to obtain our benchmarks
\be\label{eq:CI}
Y = i U_\nu  \sqrt{m_\nu}~\rho~  \sqrt{\mathcal{X_M}}^{-1},
\end{equation}
where $U_\nu$ is the lepton mixing matrix, $m_\nu$ is a diagonal matrix containing the active neutrino masses, and $\rho$ is a generic $3\times 3$ orthogonal matrix. 
For the oscillation parameters that enter $U_\nu$ and $m_\nu$, {\it i.e.}, the neutrino mixing angles, CP phase and neutrino mass splittings,
we take the values obtained in Ref.~\cite{deSalas:2020pgw} within a $3\sigma$ range, for the case of normal neutrino mass ordering. 
Moreover, we fix some of the free parameters present in $\mathcal{X_M}$ by assuming that all the relevant dimensionless scalar couplings are $\lambda_i = 0.1$, the bare masses for the dark scalars $\mu_{\eta}=\mu_{\sigma}=1$ GeV,
while $\mu_1 = 5$ TeV, and the Yukawa couplings are $Y^N=Y^f_a=0.5$.

\begin{figure}[h!]
 \centering
 \includegraphics[scale=0.75]{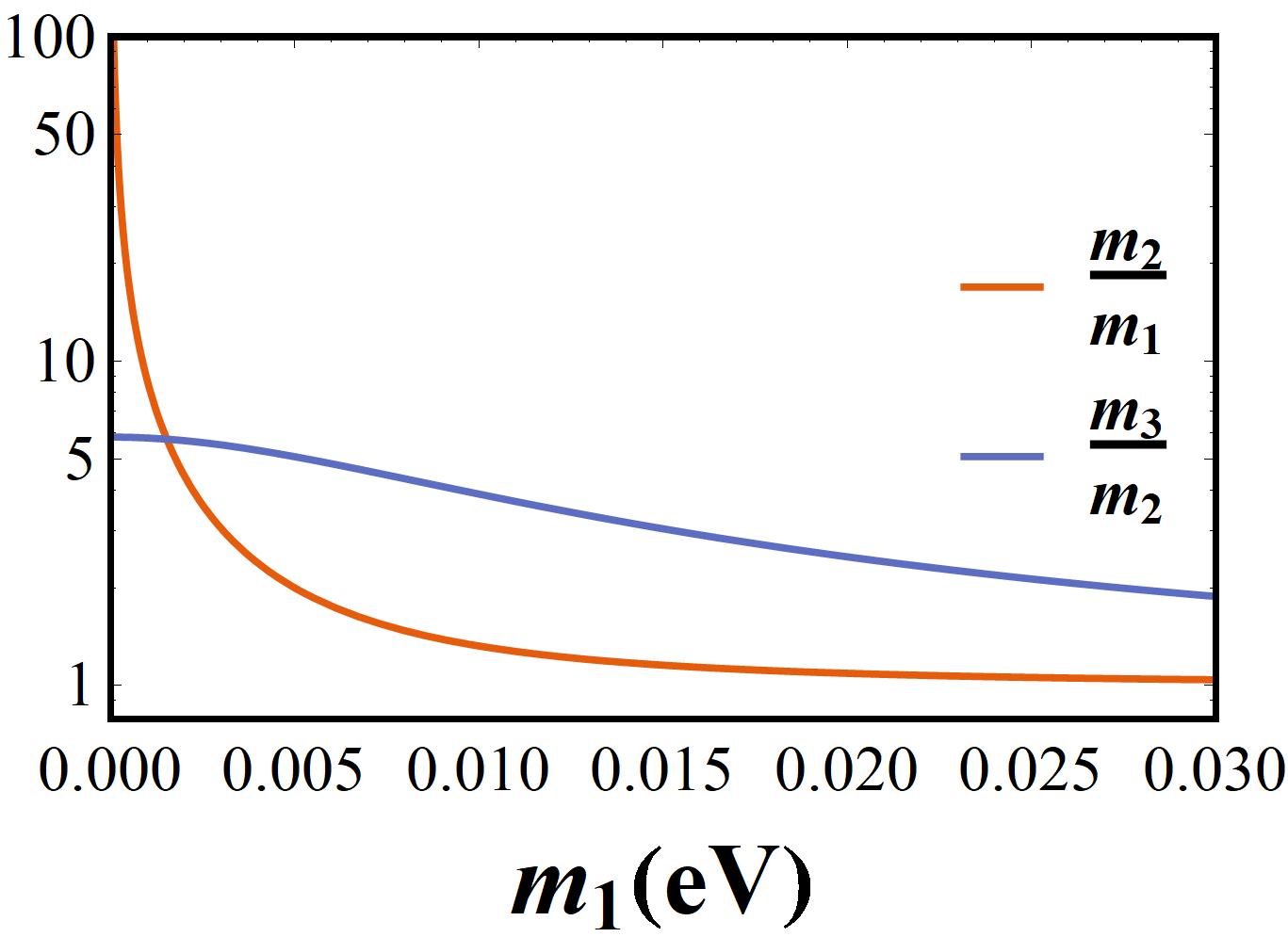}
 \caption{Neutrino mass hierarchies versus the lightest neutrino mass, $m_1$, for normal ordering. }
 \label{fig:numratios}
 \end{figure}

As for the lightest neutrino mass, $m_1$, we vary it in the range $[0.005, 0.03]$ eV.
The upper value comes from taking the upper limit on the sum of the neutrino masses: $\Sigma_i m_i < 0.12$ eV~\cite{Planck:2018vyg} together with the normal ordering assumption~\cite{deSalas:2020pgw}.
On the other hand, the lower bound comes from the radiative origin of the lighter neutrino masses, {\it i.e.}, $m_1,m_2$.
In general, we expect no strong $m_2/m_1$ hierarchy or, if present, such a hierarchy would be typically milder than the $m_3/m_2$ hierarchy, since $m_3$ has a tree-level origin. 
This is seen from Fig. \ref{fig:numratios}, where we show the $m_3/m_2$ and $m_2/m_1$ mass ratios versus $m_1$ for normal ordering.
One sees that this is indeed the case, as long as $m_1\geq 0.005$ eV.

\begin{figure}[h!]
\centering
\includegraphics[scale=.6]{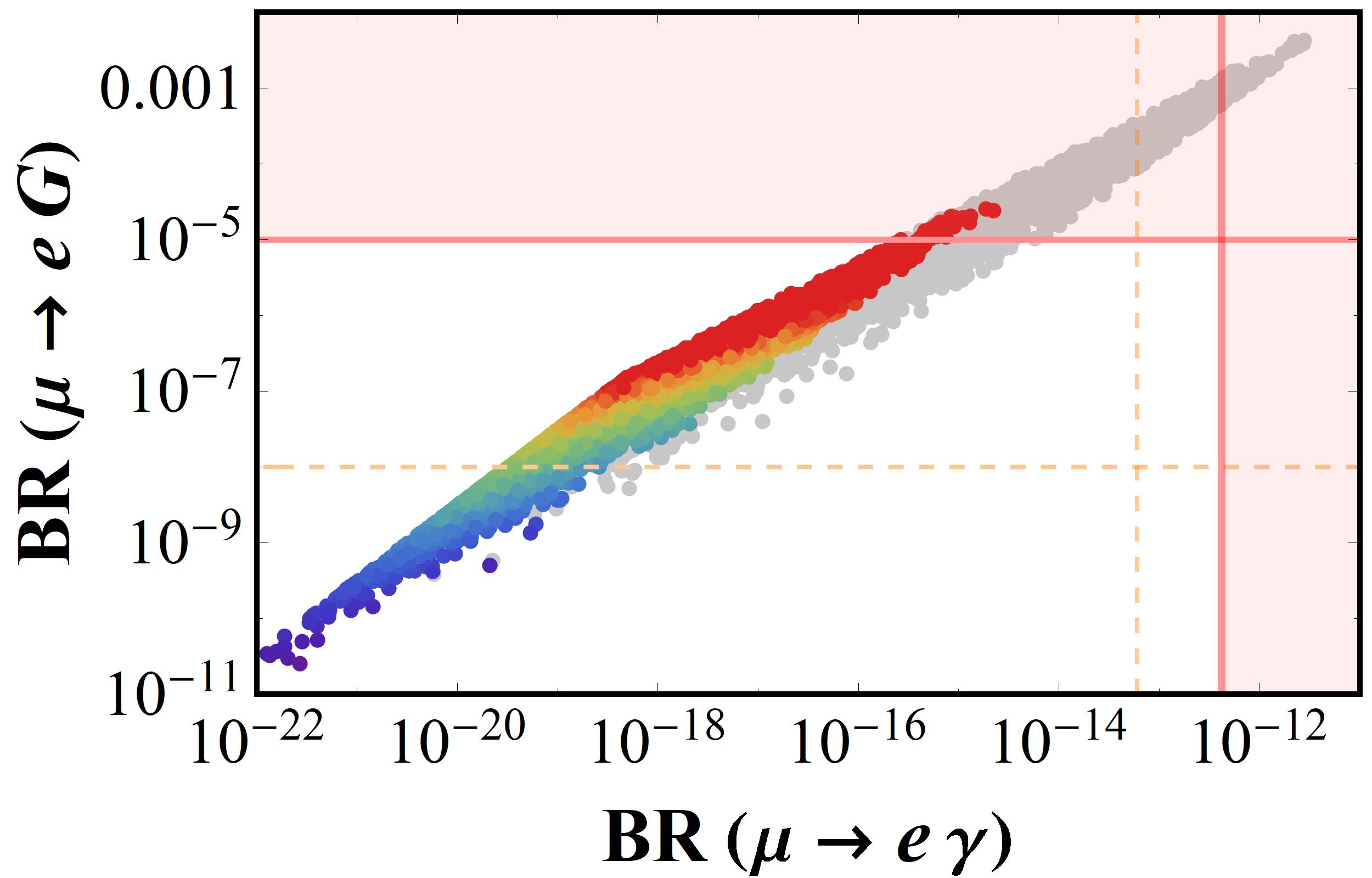}\hspace{0.3cm}
\includegraphics[scale=.6]{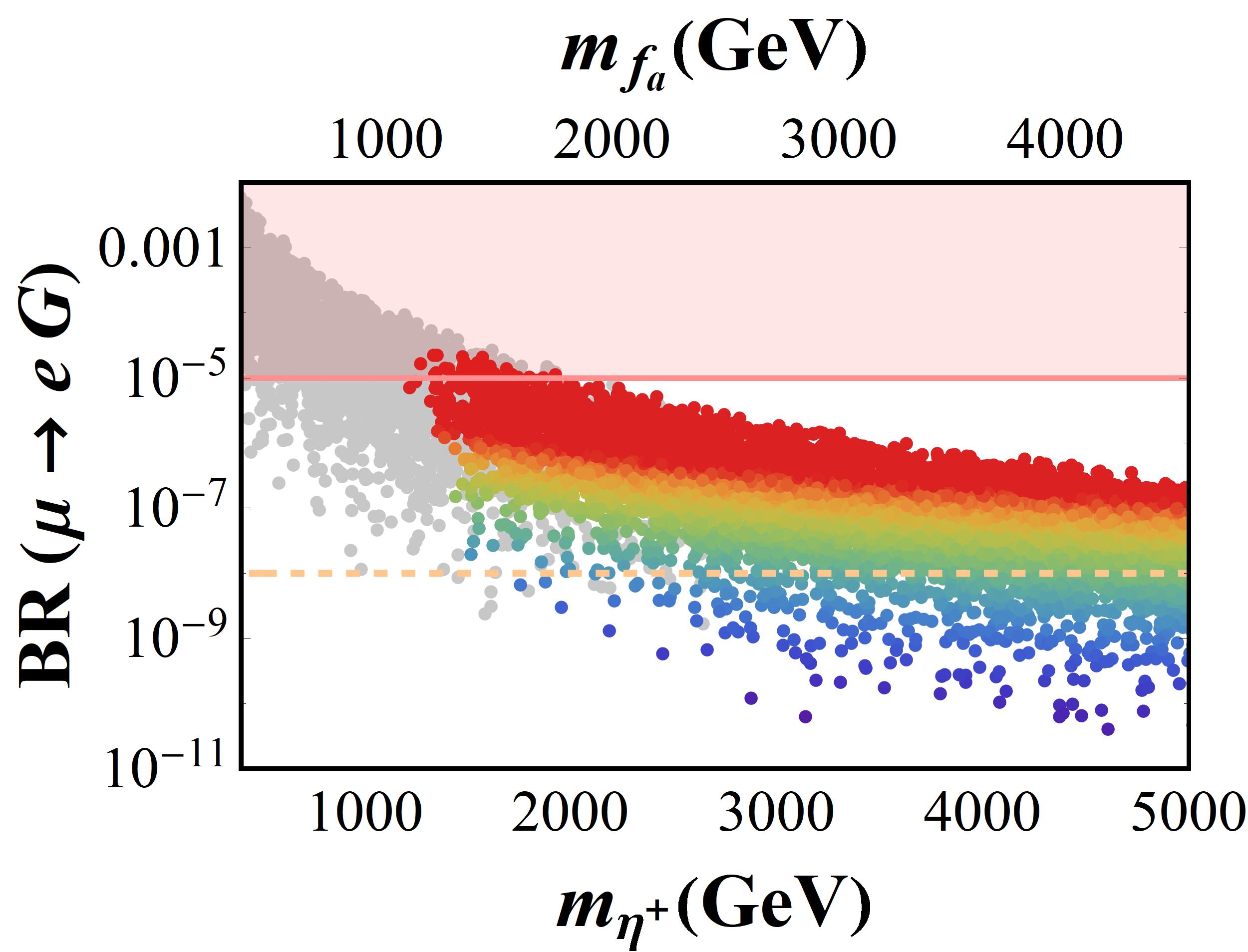}\hspace{0.1cm}
\includegraphics[scale=.6]{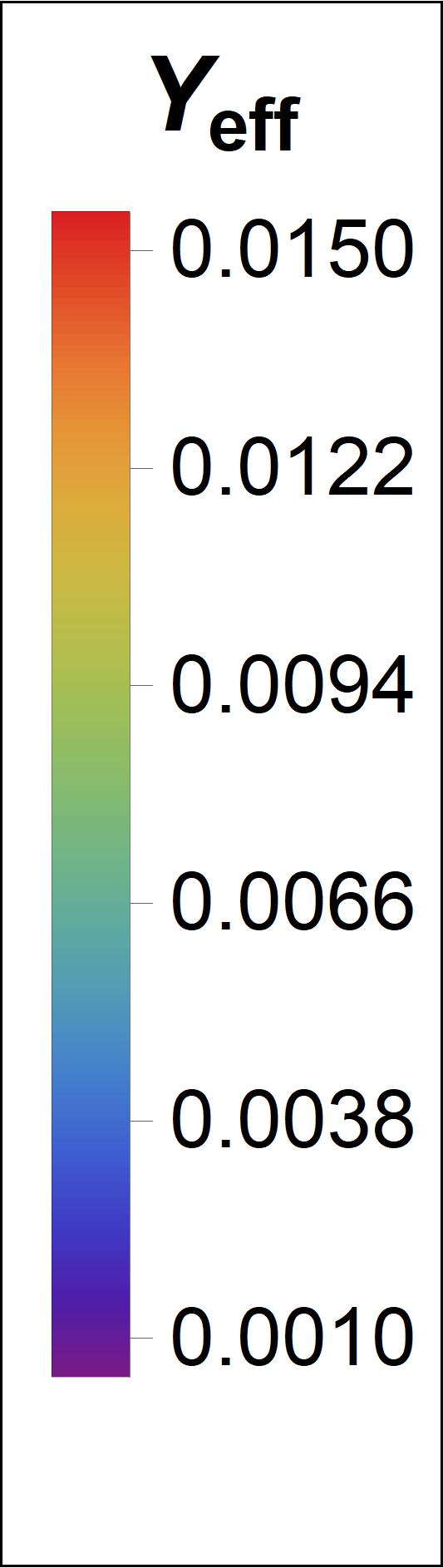}
\caption{ Left: branching ratios for $\mu\to e G$ vs $\mu\to e \gamma$. Right: branching ratio for $\mu\to e G$ vs the masses of the dark mediators, $m_{\eta^+}, m_{f_a}$, given as functions of $v_\varphi$.
The shaded red bands are excluded by the current limits on the branching ratios $BR(\mu\to e G)$ and $BR(\mu\to e \gamma)$, whereas the gray points are excluded by the limits on the diagonal couplings
  ($\Sigma^{G\,ee}, \Sigma^{G\,\mu\mu}$). The dashed orange lines are the expected experimental sensitivities, as discussed in the text. The colour bar denotes the effective Yukawa defined as
  $Y_{\mathrm{eff}} = \sqrt{|\sum_a Y^\eta_{e a}Y^{\eta *}_{\mu a}|}$. In this scenario, $v_\varphi \in [1,15]$ TeV and $v_\Phi \in [1,277]$ MeV.} 
\label{fig:cLFV1}
\end{figure}

 In what follows, we consider two scenarios.
 In the first, the induced VEV $v_\Phi$, defined in Eq. (\ref{eq:vevPhi}), varies with varying $v_\varphi\equiv v_{\varphi_{1,2,3}}$, which describes the $B-L$-breaking scale,
 whereas in the second case, $v_\Phi$ remains constant although $v_\varphi$ varies.   
 The other parameters -- such as the scalar potential couplings ($\lambda_i, \mu_i$) and the Yukawa couplings ($Y^f_a$) -- are assumed constant throughout the scans, as explained below Eq. (\ref{eq:CI}). This choice helps isolating the impact of $v_\varphi$ on the dark mediator masses and the cLFV branching ratios.
 Our results for the first scenario are presented in Fig. \ref{fig:cLFV1}.
 In this case, $v_\varphi$ varies randomly between $1$ and $15$ TeV, and the induced VEV $v_\Phi$ varies with $v_\varphi$, according to Eq. (\ref{eq:vevPhi}), which leads to $v_\Phi$ values roughly between $1$ and $277$ MeV. 
 The red lines at $10^{-5}$ (horizontal) and $4.2\times 10^{-13}$ (vertical) indicate the experimental bounds for the branching ratios of $\mu\to e G$  \cite{Jodidio:1986mz,Hirsch:2009ee,TWIST:2014ymv} and $\mu \to e \gamma$ \cite{MEG:2016leq}, respectively, so that the points within the shaded red areas are excluded.
 Moreover, the gray points are excluded by the astrophysical constraints on the flavour-diagonal Goldstone couplings $\Sigma^{G\,ee}$ and/or $\Sigma^{G\,\mu\mu}$.
 The dashed orange lines represent future experimental sensitivities for both processes, {\it i.e.}, $BR(\mu\to e \gamma)\lesssim 6\times 10^{-14}$ from the MEG II experiment \cite{MEGII:2021fah} and
 $BR(\mu\to e G)\lesssim\mathcal{O} (10^{-8})$ from the COMET experiment \cite{COMET:2018auw}, as derived in Ref. \cite{Xing:2022rob}.
 The remaining points are coloured according to the ``size'' of the corresponding Yukawas, defined as $Y_{\mathrm{eff}} = \sqrt{|\sum_a Y^\eta_{e a}Y^{\eta *}_{\mu a}|}$, as shown in the colour bar on the right.
  One sees from the left panel of Fig.~\ref{fig:cLFV1} that $\mu\to e G$ is a very promising cLFV channel.
   Indeed, in order to comply with the existing $\mu\to e G$ limit, $BR(\mu\to e \gamma)$ must be in the $10^{-14}$ range or less, which will be somewhat challenging for the upcoming round of experiments.
 As can be seen from the right panel, this implies lower bounds for the dark mediator masses.

\begin{figure}[h]
\centering
\includegraphics[scale=.6]{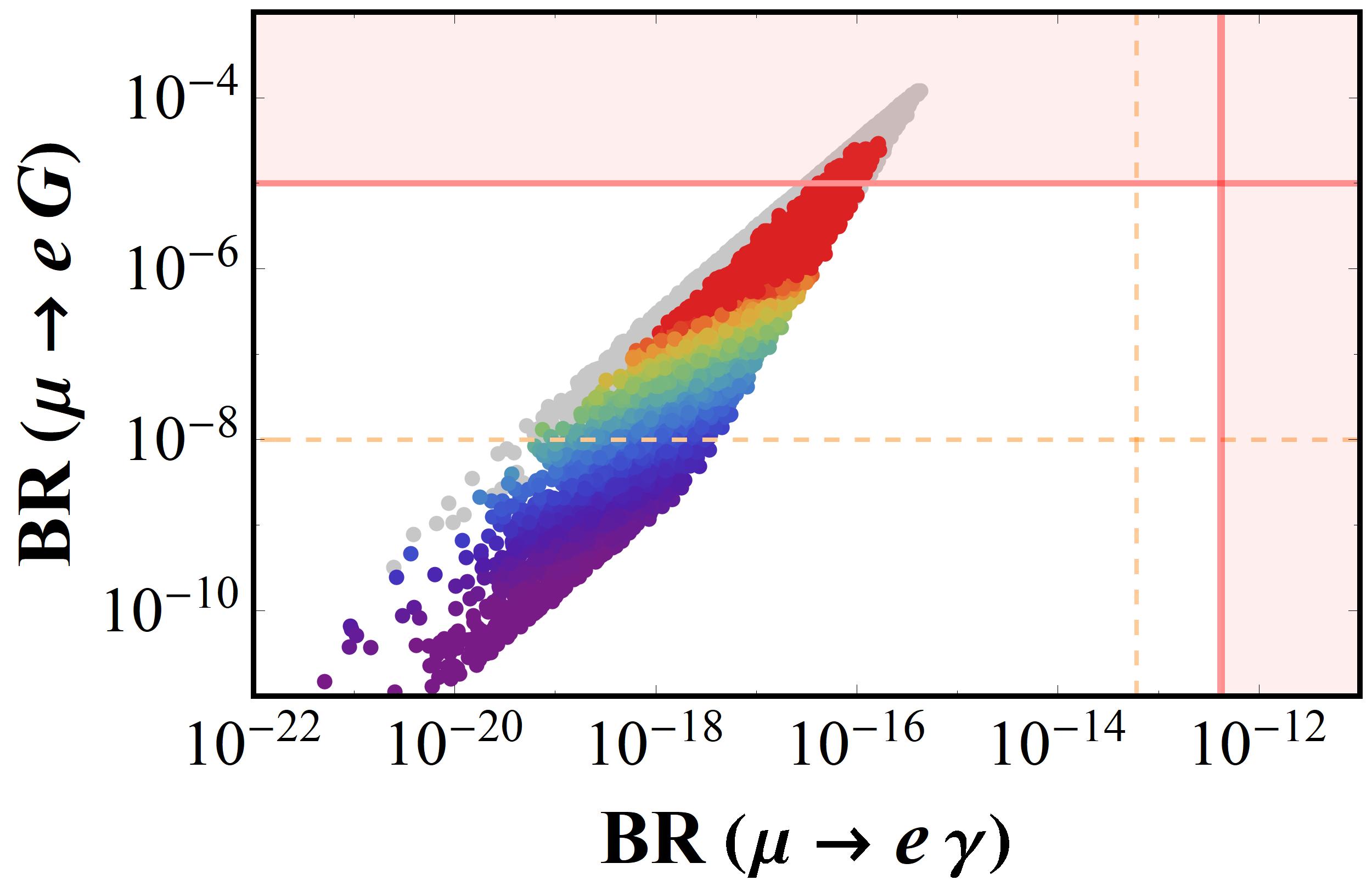}\hspace{0.3cm}
\includegraphics[scale=.6]{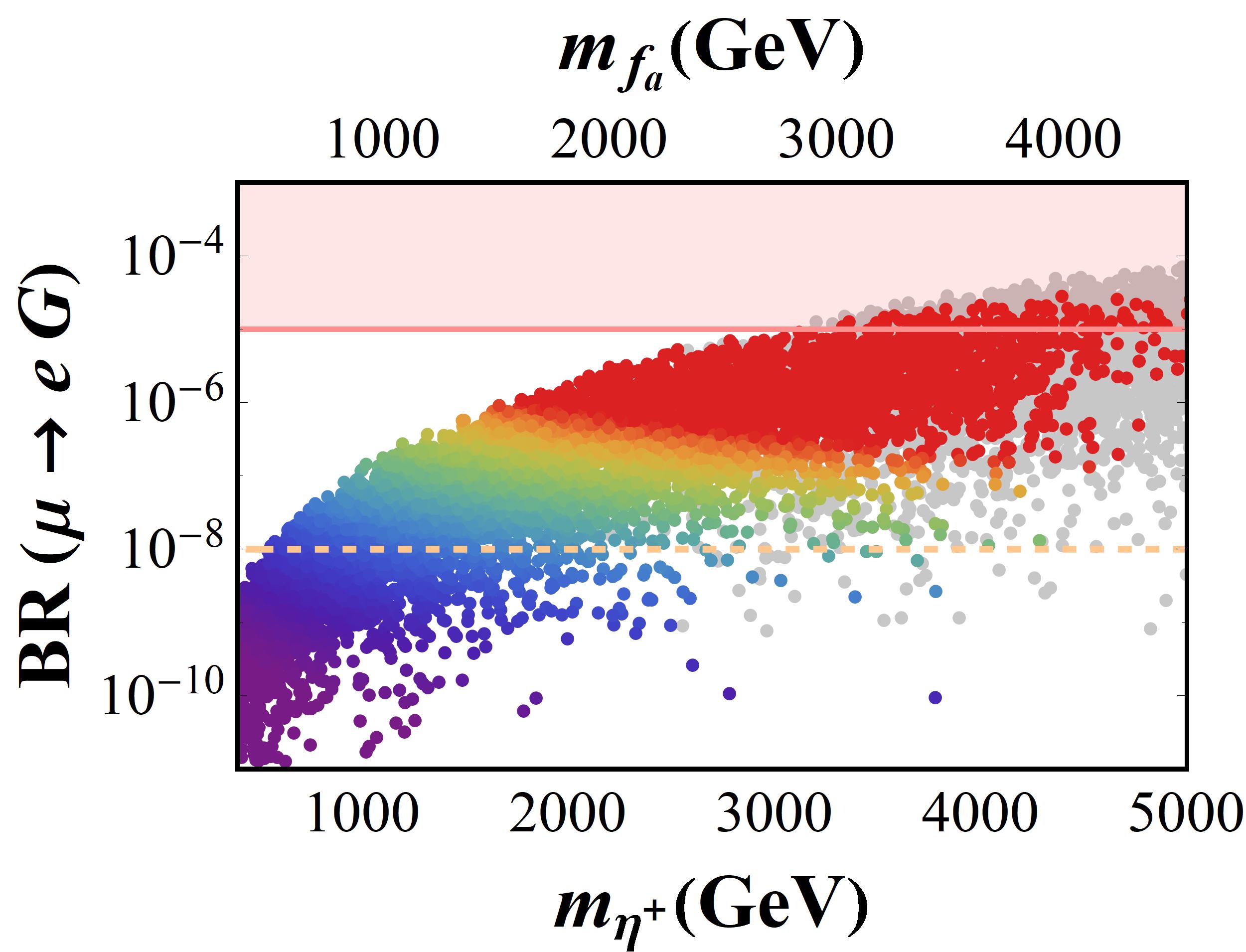}\hspace{0.1cm}
\includegraphics[scale=.6]{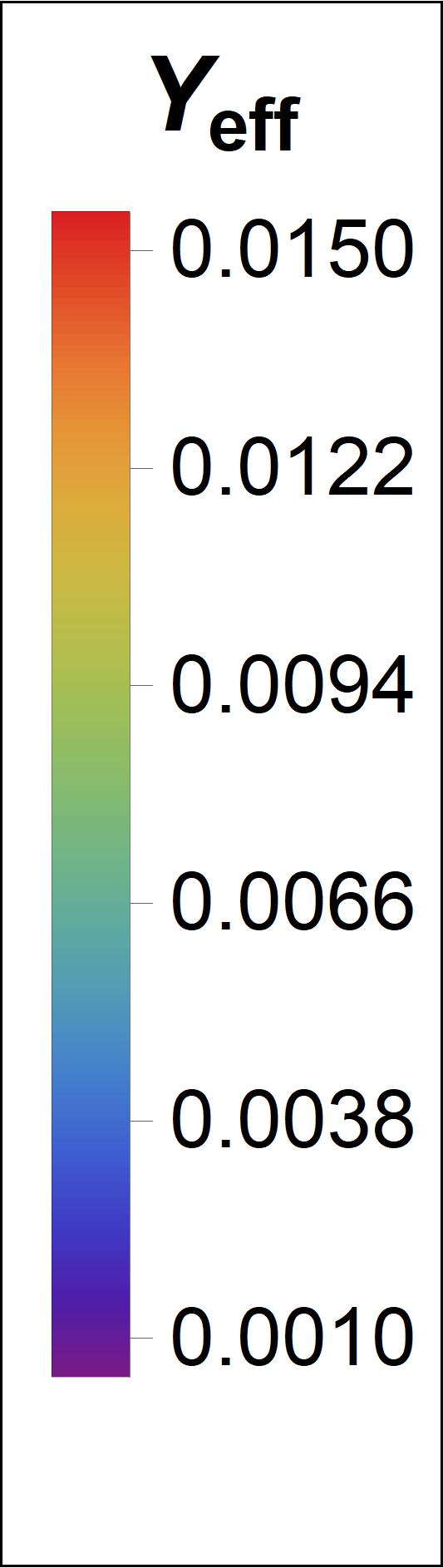}
\caption{Left: branching ratios for $\mu\to e G$ vs $\mu\to e \gamma$. Right: branching ratio for $\mu\to e G$ vs the masses of the dark mediators, which are both functions of $v_\varphi$. Same colour code as in Fig. \ref{fig:cLFV1}. The relevant Yukawas follow the colour bar on the right, with $Y_{\mathrm{eff}} = \sqrt{|\sum_a Y^\eta_{e a}Y^{\eta *}_{\mu a}|}$. In this scenario, we assume $v_\varphi \in [1,15]$ TeV, whereas the induced VEV is fixed: $v_\Phi = 50$ MeV. }
\label{fig:cLFV2}
\end{figure}

We now turn to a scenario in which the induced VEV $v_\Phi$ is a constant, {\it i.e.}, $v_\Phi = 50$ MeV, but $v_\varphi$ is allowed to vary,
as previously, between $1$ and $15$ TeV. The plots in Fig. \ref{fig:cLFV2} show very different behaviours when compared to those in the first scenario. 
Here, the branching ratios increase as the mediator masses, $m_{\eta^+}, m_{f_a}$, increase, whereas the opposite happens in the first scenario.

To understand this, notice first that the branching ratios in Eqs. (\ref{eq:mu2eg2}) and (\ref{eq:mu2eG2}) vary roughly as $Y_{\mathrm{eff}}^4( m_\mu /v_\varphi)^4 $ and $ Y_{\mathrm{eff}}^4 (m_\mu/v_\varphi)^2$ \footnote{In the region of interest, the dark mediator masses, $m_{f_{1,2}}$ (assumed to be degenerate) and $m_{\eta^+}$, vary linearly with $v_\varphi$ in such a way that the loop functions $F_{1,2}(m_{f_a}/m_{\eta^+})$ are approximately constants.}.
The entries of $Y_{\mathrm{eff}}=\sqrt{|\sum_a Y^\eta_{e a}Y^{\eta *}_{\mu a}|}$ contribute to (scotogenic) solar neutrino mass generation which depend on $v_\Phi^2$; see Fig. \ref{fig:massdiags}. 
Thus, once we impose neutrino oscillation data, via Eqs. (\ref{eq:redef}) and (\ref{eq:CI}), $Y_{\mathrm{eff}}$ can also be described as a function of the induced VEV $v_\Phi$.
In the first scenario (Fig. \ref{fig:cLFV1}), as $v_\Phi$ increases with $v_\varphi$, following Eq. (\ref{eq:vevPhi}), the scotogenic loop functions $\mathcal{M}_c$ in Eq. (\ref{scotomass}) also increase.
Consequently, to ``neutralise'' this increase in $\mathcal{M}_c$, so as to satisfy neutrino oscillation data, the relevant Yukawas should decrease accordingly.
Therefore, since both $Y_{\mathrm{eff}}$ and $(m_\mu/v_\varphi)$ decrease with increasing $v_\varphi$, the branching ratios also decrease, as shown in Fig. \ref{fig:cLFV1}.
In contrast, in the scenario where $v_\Phi$ is constant, the scotogenic loop functions $\mathcal{M}_c$ decrease with increasing $v_\varphi$, and, in turn, the Yukawas increase to satisfy oscillation data.
Therefore, while $(m_\mu /v_\varphi)$ decreases, $Y_{\mathrm{eff}}$ ends up increasing, with $v_\varphi$. Nevertheless, it turns out that the dependence of $Y_{\mathrm{eff}}$ on $v_\varphi$ is stronger than linear so that the branching ratios,
given as functions of $Y_{\mathrm{eff}}^4( m_\mu /v_\varphi)^4 $ and $ Y_{\mathrm{eff}}^4 (m_\mu/v_\varphi)^2$, end up increasing with increasing $v_\varphi$, as shown in Fig. \ref{fig:cLFV2}.\\

In short, we have seen that our model naturally leads to detectable cLFV rates involving Goldstone boson emission.
  Although we have mainly focused on the implications of Goldstone boson emission in cLFV processes, there is a rich phenomenology associated to Goldstone boson processes that can affect other sectors.
  For instance, the SM Higgs boson can decay invisibly $h\equiv S_1 \to GG$~\cite{Joshipura:1992hp,Eboli:1994bm} and also the effective number of relativistic degrees of freedom could be affected. 
  Our model is elastic enough to satisfy constraints such as the invisible Higgs decays~\cite{ATLAS:2019cid,CMS:2018yfx} by adequately choosing the Higgs potential parameters. For a recent discussion on similar models see~\cite{Fontes:2019uld}.
  Likewise, by properly choosing the Higgs potential parameters, $Z^\prime$ mass and couplings one can suppress excessive contributions to the effective number of relativistic degrees of freedom~\cite{Alvarado:2021fbw}
  so that it can fulfill the observational constraint~\cite{Planck:2018vyg}.
  Therefore, the model offers sufficient flexibility to accommodate constraints arising from the Goldstone boson interactions without compromising the promising prospects of detecting cLFV through Goldstone boson emission.
  A detailed analysis of the parameter space falls outside the scope of the present work.

\section{Collider signatures}
\label{subsec:collider}

\subsection{$p p  \to Z^{\prime} \to N N$ Drell-Yann Production}

In the simplest scoto-seesaw model~\cite{Rojas:2018wym,Mandal:2021yph}, the neutrino mass mediators are unlikely to be produced at colliders.
For example, the atmospheric neutrino mass mediator, {\it i.e.}, the heavy neutral lepton $N$ (right-handed neutrino), is inaccessible through the Drell-Yan mechanism.
Indeed, $N$ is typically very heavy. Even if (artificially) taken to lie in the TeV scale, its production would be suppressed by a tiny doublet-singlet $\nu-N$ mixing. 

In the present case, however, both the neutral lepton $N$ and the vector boson $Z^\prime$ can lie in the TeV-scale.
The existence of the $Z^\prime$ provides a Drell-Yan pair-production portal for the heavy neutrino. 
Its subsequent decays into the SM states could potentially lead to verifiable signals.
Due to the enhanced gauge charge (5) of the right-handed fermion $N$, its $Z^\prime$ phenomenology can differ from that of other $U(1)_{B-L}$ models.

We remark that the neutrino mass suppression mechanism of our scoto-seesaw model involving the induced VEV of the leptophilic doublet $\Phi$, Figs.~\ref{fig:massdiags},
requires a rich scalar sector compared to a vanilla $U(1)_{B-L}$ model. 
We have an extended visible scalar sector with four more CP-even scalars apart from the Higgs boson. 
Among them, the scalar predominantly coming from the leptophilic doublet $\Phi$ is relatively heavy and beyond the LHC reach, while the others are lighter, possibly around the TeV scale.
Moreover, there are two massive CP-odd scalars. While one of them is heavy, as it comes mostly from $\Phi$, the other can potentially lie at the TeV scale. 

The $p p \to N N$ collision process can also proceed through the exchange of neutral scalars.   
These couple to the initial-state quarks and final-state neutral fermion pair through the mixing of the Higgs boson with other CP even scalars.
Nonetheless, for simplicity, we assume that this mixing is not significant so that such decays ($ S_i\to NN$) are not kinematically allowed,
and these contributions can be neglected.  
In Table~\ref{Tab2}, we give estimates of the expected $Z^\prime$ mediated production cross section. 
\begin{table}[h!]
\begin{center}
\begin{tabular}{|  c|| c |}
  \hline 
   \centering{Benchmark Points}  &  \,\,$\sigma (p p \to N N)$ (in fb)      \\
\hline \hline
  BP-I: $m_{Z^{\prime}}$  = 1.5 TeV, $m_N = 708$~GeV\,\,
($Z^{\prime}$ decays to dark sector are allowed) & 0.489 \\\hline
 \,\,BP-II: $m_{Z^{\prime}}$  = 1.5 TeV, $m_N = 708$~GeV\,\,
 ($Z^{\prime}$ decay to dark sector negligibly small) \,\, & 4.76  \\
 \hline
BP-III: $m_{Z^{\prime}}$  = 2.0 TeV, $m_N = 708$~GeV \,\,&   5.14   \\
\hline
 BP-IV: $m_{Z^{\prime}}$  = 2.5 TeV, $m_N = 708$~GeV\,\, &  1.44 \\
 \hline
 BP-V: $m_{Z^{\prime}}$  = 3.0 TeV, $m_N = 708$~GeV\,\,  &  0.415 \\
\hline \hline
		
  \end{tabular}
\end{center}
\caption{
  Drell-Yan neutral lepton pair production through the $Z^{\prime}$ portal.
  The cross section is computed with $\lambda_{ii} \sim 0.1, \lambda_{ij} \sim 0.01, Y_i \sim 0.5, g^\prime = 0.1 $.
  The dominant contribution is through the $Z^{\prime}$ mediated process $\sigma (p p \to Z^{\prime} ) \times BR(Z^{\prime} \to NN)$.  }
 \label{Tab2} 
\end{table}

Note that the neutral fermion mass is taken around 700 GeV, allowing the $Z^{\prime}$ boson to decay to it in on shell. 
For a $Z^{\prime}$ mass of 1.5 TeV, the branching ratio to $Z^{\prime} \to NN$ is relatively low, while it peaks near 2 TeV $Z^{\prime}$ mass, and saturates at around $25\%$. 
Therefore, $\sigma(pp \to Z^{\prime}) \times BR (Z^{\prime} \to NN)$ increases initially up to 2 TeV, and then it gradually decreases with increasing $Z^{\prime}$ mass.
Notice that the heavy neutrino $N$ will decay to SM states, such as leptons, through $\nu-N$ mixing in the electroweak currents.
This leads to leptonic final-state signatures such as dileptons plus missing energy\footnote{
  Dark sector fermions can also be produced through Drell-Yann processes
  like $\sigma (p p \to f_i f_i)$, with less striking signatures.  }. 
It is not in our scope here to present a dedicated numerical simulation of these signatures. A detailed collider analysis of the $p p \to Z^{\prime} \to NN $ mode in a slightly different scenario is performed in Ref.~\cite{Arun:2022ecj}.

\subsection{$Z^{\prime}$ Phenomenology: Dilepton search} 

ATLAS and CMS experiments are looking for a heavy BSM gauge boson in different channels.
The leading LHC production mode for a $Z^{\prime}$ with couplings to quarks (as is the case for all $U(1)_{B-L}$ models) is $q\bar{q}$ fusion.  
In our model, we have many decay modes of $Z^{\prime} \to S_i A_j$ and $Z^{\prime} \to S_i Z$, where $S_i$ and $A_j$ are the CP-even and CP-odd scalars, 
in both visible and dark scalar ($s_i, a_i$) final states. 
As a consequence, our $Z^{\prime}$ decay rates to the SM channels can be relatively small. 

\begin{figure}[h!]
\centering
\includegraphics[scale=0.28]{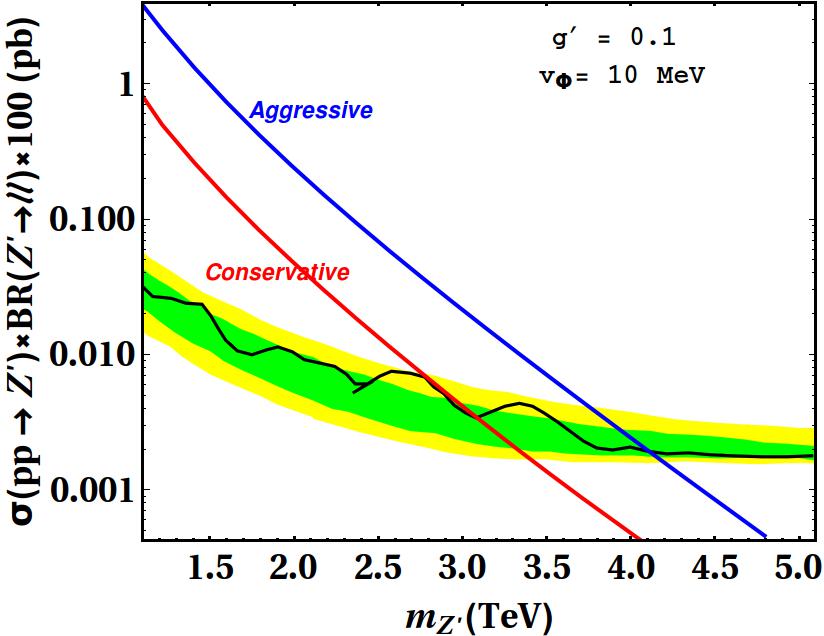}~~
\includegraphics[scale=0.26]{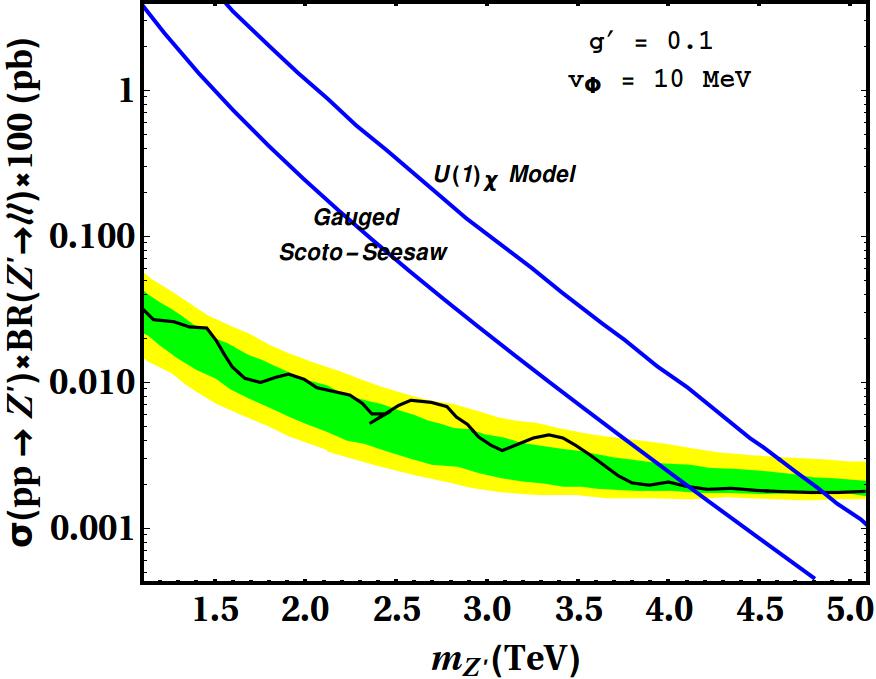}~~
\caption{
  The left panel gives the expected $\sigma \times BR$ of the dilepton signal in our dynamical scoto-seesaw model, obtained when all exotic channels are accessible
  (red line) or kinematically forbidden (blue line); see text.
  The black line represents the dilepton channel measurement from ATLAS~\cite{ATLAS:2019erb} (with similar results also from CMS~\cite{CMS:2018ipm}),
  while the green and yellow bands give $1 \sigma$ and $2 \sigma$ regions, obtained from the LHC data repository HEPData~\cite{Maguire:2017ypu}. 
  Exclusion limits on the $Z^{\prime}$ mass within these two extreme benchmarks are readily obtained from the blue and red curves. 
  The right panel compares the aggressive constraints (blue line) obtained in our model and in a reference $\chi$ model.
}
\label{fig:collider-dilepton}
\end{figure}

Despite such a smaller branching ratio to the SM channels, the $Z^{\prime}$ decay to a pair of charged leptons places strong constraints on the $Z^{\prime}$ mass and coupling;
see Fig.~\ref{fig:collider-dilepton}. 
 The exclusion bounds on the parameters of our model at the LHC are derived by comparing 
our theoretical predictions with the experimental limits obtained from the non-observation of any BSM excess in a particular mode.  
Here, we fix the gauge coupling at a value $g^{\prime} = 0.1$ and compare the theoretical cross section $\sigma (p p \to Z^{\prime}) \times BR (Z^{\prime} \to  \ell \ell)$
 in our model with the upper bound obtained from the LHC dilepton search. This way we provide limits on the $Z^{\prime}$ mass, $M_{Z^{\prime}}$.  
 The $Z^{\prime}$ limit in the sequential standard model (SSM) obtained by CMS~\cite{CMS:2018ipm} excludes any new boson below 4.4 TeV. 
 In a similar $U(1)_{B-L}$ model with $5, -4, -4$ gauged lepton number charges~\cite{Choudhury:2020cpm},
 in the presence of higher dimensional operators, the lower bound on $Z^{\prime}$ hovers around $m_{Z^{\prime}} \approx 2.5-3$~TeV.    

Considering the left plot, the blue line gives the strongest constraint, excluding any $Z^{\prime}$ mass up to $m_{Z^{\prime}} \sim 3.9$~TeV. 
  This is the case where all scalar quartic couplings, except for $\lambda_H$, chosen to fit the SM Higgs boson mass, are taken to be large $\lambda_i \sim 0.75$,
  leading to a heavier scalar spectrum which restricts $Z^{\prime}$ decays to BSM scalar modes. 
  Likewise all BSM fermion Yukawa couplings are taken large $Y_a = 1.5$ so that also the $N, f_a$ fermions are too heavy to be produced by $Z^{\prime}$ decays,
  leading to a smaller branching ratio for $Z^\prime$ to charged leptons, {\it i.e.}, $ BR (Z^{\prime} \to \ell_i \ell_i) \sim 15\%$.
  Our second benchmark corresponding to the red line leads to weaker constraints.
  In this case, the scalar quartics take values $\lambda_i \sim 0.10$, making the scalar sector relatively light so that $Z^{\prime}$ can decay to some dark scalar-pseudoscalar pair. 
  Even in this limit, $Z^{\prime}$ decays to visible sector scalars are not favored, mainly due to phase space.
  As for the BSM Yukawa couplings, we assume $Y_a = 0.5$, which opens up decay modes like $Z^{\prime } \to NN, f_a f_a$ with BRs in the order of 30\% and 20\% each, respectively. 
 These new decay modes suppress the dilepton BR to $ BR (Z^{\prime} \to \ell_i \ell_i) \sim 2-4 \%$, leading to a weaker limit on the $Z^{\prime}$ mass, $m_{Z^{\prime}} \ge 2.7~$TeV.
 Finally, in the right plot, we compare our ``aggressive'' model prediction to that of the $U(1)_\chi$ model, an $E6$-motivated Grand Unification construction \cite{London:1986dk}. We observe that the lower bound for the $Z^\prime$ mass of the $U(1)_\chi$ model lies around $4.8$ TeV, roughly $1$ TeV above the limit for $m_{Z^\prime}$ in our model.
 For details regarding the collider aspects of $Z^{\prime}$ phenomenology, we use Ref.~\cite{Choudhury:2020cpm, Deka:2021koh}. 

\section{Dark matter} 
\label{subsec:dm}

In our gauged scoto-seesaw model, dark matter can be either the lightest ``dark'' scalar or fermion, stabilized through the residual matter parity symmetry defined in Eq.~(\ref{eq:MP}).
Before embarking on our dark matter discussion we note that, for the particular benchmarks adopted in the previous sections, the dark matter candidate is always a scalar field.
Radiative exchange of the ``dark'' particles generates the solar neutrino mass scale, as illustrated in the right panel of Fig.~\ref{fig:massdiags}. 
The dark sector includes scalar dark matter candidates that are mixtures of neutral scalars of the $SU(2)_L$ doublet $\eta$ and the singlet $\sigma$, plus the charged scalar $\eta^{\pm}$. 
 On the other hand, the fermionic dark sector contains two neutral Majorana fermions $f_a$.
In addition to the standard Higgs portal mediated DM annihilation, we can have a novel $Z^\prime$ mediated annihilation, along with some new scalar mediated annihilation.

\begin{itemize}
\item {\bf Scalar dark matter}
  
  The scalar dark matter sector of our scoto-seesaw model includes, besides the doublet $\eta$, an additional singlet $\sigma$.
  The neutral components of the singlet $S_{\sigma}, A_{\sigma}$ and doublet $S_{\eta}, A_{\eta}$ mix so as to produce the CP-even ($s_1, s_2$) and CP-odd ($a_1, a_2$) DM candidates, respectively.
 (In contrast, the dark scalar sector of the simplest scoto-seesaw scenario, analysed in Ref.~\cite{Rojas:2018wym,Mandal:2021yph}, consists of a single scalar doublet $\eta$.) 
As the doublet-singlet mixing is small, see Eq.~(\ref{eq:mix}), two of the scalar DM candidates, one CP-even $s_1$ and one CP-odd $a_1$, are mainly doublet DM candidates.
The neutral components of the doublet $\eta$ resonantly annihilate through the Higgs portal.  
In our setup, we can have additional resonant annihilation through non-dark CP-even Higgs bosons ($S \equiv S_{\Phi}, S_{\varphi_i}$).
This may lead to enhanced annihilation and underabundant relic at different DM masses $m_{\rm DM} \sim m_S/2$.  

In addition to the $\eta$ mediated t-channel annihilation to SM states $\ell \ell, WW, ZZ, hh$, present in the simplest scoto-seesaw (Ref.~\cite{Rojas:2018wym,Mandal:2021yph}),
in our new model, there are other t-channel contributions to scalar DM annihilation, which can enhance the annihilation cross section,
leading to more parameter space with an underabundant relic.

The dark singlet $\sigma$ present in our model annihilates through similar channels as the doublet-like dark matter, though the relative dominance of different contributions
will depend on the mixing of the CP-even scalars $h,S_i$.  
For example, singlet $s_2$ and $a_2$ annihilation through quartic couplings like $SSVV$ with $S \equiv s_2, a_2$  and $\rm V \equiv W^{\pm}, Z$ are absent,
whereas same vertices with $V \equiv Z^{\prime} $ and the corresponding annihilation channel will be present.

Moreover, the scalar dark sector phenomenology of our model will be modified due to the presence of a $Z^{\prime}$ portal.
This happens because, if CP-even ($s_{1,2}$) and CP-odd ($a_{1,2}$) dark scalars are almost degenerate, co-annihilation happens through a $Z^{\prime}$ mediated s-channel process.
This coannihilation adds up to the existing DM annihilation modes, opening up new parameter space consistent with measured relic density.  
 
\item {\bf Fermionic singlet dark matter} 

  In our gauged scoto-seesaw model, two fermionic dark matter candidates $f_a$ mediate the loop-induced solar mass scale,
  instead of a single dark fermion in the simplest scoto-seesaw~\cite{Rojas:2018wym,Mandal:2021yph}. 
These dark fermions $f_a$ have a new $Z^\prime$ portal to annihilate.
In contrast to the simplest non-gauged scoto-seesaw model in Ref.~\cite{Rojas:2018wym,Mandal:2021yph},
this brings a significant enhancement in the DM annihilation, including a resonant dip in the relic abundance at $m_{\rm DM} \sim m_{Z^{\prime}}/2$. 

Moreover, in the simplest scoto-seesaw mechanism, the dark fermion mass term is a bare mass term.
As a result, there is no Higgs-mediated s-channel fermionic dark matter annihilation channel. 
In contrast, in our gauged scoto-seesaw model there is a singlet $\varphi_2$ that provides mass to the dark Majorana fermions $f_a$,
and can act as a portal of DM annihilation, creating a resonant dip in the relic density. \\

\item {\bf Constraints and dark matter detection}
  
  In the standard scoto-seesaw (Ref.~\cite{Rojas:2018wym,Mandal:2021yph}), light doublet-like dark matter is constrained from the LEP Z decay measurement.
  In vanilla scotogenic models, including the original one~\cite{Ma:2006km} and its simplest triplet extension~\cite{Hirsch:2013ola},
  nearly degenerate neutral CP-even and CP-odd scalars from $\eta$ is typically assumed in order to implement scotogenic neutrino mass generation.  
     In our present model, however, instead of relying on the quasi-degeneracy of the scalar mediators, the smallness of neutrino masses follows mainly from their
    dependence on the (small) induced VEV $v_\Phi$, Fig.~\ref{fig:massdiags}. 
    Consequently, here neutrinos can be light enough even when one has a light CP-even DM with a heavier CP-odd counterpart, or vice versa, avoiding constraints from the $Z \to S_{\eta} A_{\eta}$ decay.
  Therefore, in our gauged scoto-seesaw a light doublet-like scalar dark matter is not as constrained as in  vanilla scotogenic
  models~\cite{Ma:2006km,Hirsch:2013ola} or the simplest scoto-seesaw scenario~\cite{Rojas:2018wym,Mandal:2021yph}.
 The LEP constraint can be completely avoided in our proposed scheme. 
 
 Our dynamical scoto-seesaw model can have light dark scalars, free from constraints that hold for a generic doublet-only dark sector,
 allowing it to harbor a viable light DM candidate. 
 The latter can be probed through dedicated experiments like CDMS-lite and CRESST, along with existing experiments, such as LZ, Xenon-nT, and PandaX. 

 Note that in our model we have DM coannihilation through a $Z^{\prime}$ portal, the DM relic density being fixed by $Z^{\prime}$ related parameters and dark sector mass degeneracy.
 Concerning direct detection, the scalar DM-nucleon scattering dominantly happens through the Higgs portal, as in generic scotogenic models. 
 This relaxes the inter-dependence between the DM annihilation and detection, as they are not controlled by the same interaction.

In the simplest scoto-seesaw scenario, for instance, there is no mediator for fermionic DM detection through DM-nucleon scattering at the tree level.
 This results in a possible loop-suppressed scattering cross section.  
The presence of DM-nucleon scattering mediated at tree level by $Z^{\prime}$ provides a significant advantage to our gauged scoto-seesaw model in this regard,
 as it enhances the possibility of direct detection of fermionic DM.
 The fermionic DM-nucleon scattering cross section then may reach the current sensitivity of the DM direct detection experiments.

\end{itemize}

\section{Conclusions} 
\label{sec:conclusions}

We have proposed a scheme where the scoto-seesaw mechanism has a dynamical origin, associated to a gauged $B-L$ symmetry.
``Dark'' states mediate solar neutrino mass generation radiatively, while the atmospheric scale arises \textit{a la seesaw}; see Fig.~\ref{fig:massdiags}.
Indeed the origin of the solar scale is \textit{scotogenic}, its radiative nature explaining the solar-to-atmospheric scale ratio.
The two dark fermions and the TeV-scale seesaw mediator carry different dynamical $B-L$ charges.
Dark matter stability follows from the residual matter parity that survives the breaking of $B-L$ gauge symmetry. 
Apart from the possibility of being tested at colliders, see Fig.~\ref{fig:collider-dilepton}, our scoto-seesaw model with gauged $B-L$ has sizeable charged lepton flavour violating phenomena.
These include also processes involving the emission of a Goldstone boson associated to an accidental global symmetry present in the theory, see Fig.~\ref{fig:cLFVdiagramsgG}.
Rate estimates for muon number violating processes are given in Figs.~\ref{fig:cLFV1} and \ref{fig:cLFV2}. They indicate that these processes lie within reach of present and upcoming searches.
Likewise, we also expect sizeable tau number violating processes.

\begin{acknowledgments}
  This work was supported by the Spanish grants PID2020-113775GB-I00~(AEI/10.13039/501100011033) and Prometeo CIPROM/2021/054 (Generalitat Valenciana).
  S.S. thanks SERB, DST, Govt. of India for a SIRE grant with number SIR/2022/000432 supporting an academic visit to the AHEP group at IFIC.
  J. V. wishes to thank Yoshi Kuno for a discussion on COMET sensitivities.
\end{acknowledgments}
\appendix
\section{Scalar sector analytical expressions}\label{app:scalar}

In this appendix, we provide in full form some of the important, but longer, analytical expressions obtained while deriving the scalar spectrum. 
The $5 \times 5$ squared mass matrix for the CP- and $M_P$-even fields $(S_H, S_\Phi, S_{\varphi_1}, S_{\varphi_2}, S_{\varphi_3})$ is given by the symmetric matrix
{\small
\bea\label{eq:MS2}
M_S^2&=&\left(
\begin{array}{ccccc}
d_{1} & \star & \star & \star & \star \\
v_H v_\Phi (\lambda_{H\Phi}+\lambda_{H\Phi}^\prime) -\frac{v_{ \varphi_2} v_{ \varphi_3} \lambda_2}{2} & d_{2} & \star & \star & \star \\
 v_H v_{ \varphi_1} \lambda_{ H\varphi_1} & v_\Phi v_{ \varphi_1} \lambda_{\Phi \varphi_1} &  d_{3} & \star & \star \\
 v_H v_{ \varphi_2} \lambda_{ H\varphi_2}-\frac{v_\Phi v_{ \varphi_3} \lambda_2}{2} & v_\Phi v_{ \varphi_2} \lambda_{\Phi \varphi_2}-\frac{v_H v_{ \varphi_3} \lambda_2}{2} & v_{ \varphi_1} v_{ \varphi_2} \lambda_{\varphi_1 \varphi_2}-\frac{v_{ \varphi_3} \mu_2}{2} &  d_{4} & \star \\
 v_H v_{ \varphi_3} \lambda_{ H\varphi_3}-\frac{v_\Phi v_{ \varphi_2} \lambda_2}{2} & v_\Phi v_{ \varphi_3} \lambda_{\Phi \varphi_3}-\frac{v_H v_{ \varphi_2} \lambda_2}{2} & v_{ \varphi_1} v_{ \varphi_3} \lambda_{\varphi_1 \varphi_3}-\frac{v_{ \varphi_2} \mu_2}{2} & v_{\varphi_2}v_{\varphi_3} \lambda_{\varphi_2\varphi_3} - \tilde{m}^2 & d_{5}\\
\end{array}\,
\right)\,,
\eea}
with
\bea\label{eq:MS2d} 
d_1&=& 2 v_H^2 \lambda_H+\frac{v_\Phi v_{ \varphi_2} v_{ \varphi_3} \lambda_2}{2 v_H}, \quad d_2 =2 v_\Phi^2 \lambda_{\Phi}+\frac{v_H v_{ \varphi_2} v_{ \varphi_3} \lambda_2}{2 v_\Phi}, \quad d_3 = 2v_{ \varphi_1}^2 \lambda_{\varphi_1}+\frac{v_{ \varphi_2} v_{ \varphi_3} \mu_2}{2 v_{ \varphi_1}}, \\
d_4 &=& 2 \lambda_{\varphi_2} v_{ \varphi_2}^2+\frac{v_{ \varphi_3}\tilde{m}^2}{v_{ \varphi_2}} , \quad d_5 =2 \lambda_{\varphi_3} v_{ \varphi_3}^2+\frac{v_{ \varphi_2}\tilde{m}^2}{ v_{ \varphi_3}}, \quad \tilde{m}^2 =\frac{v_H v_\Phi \lambda_2+v_{ \varphi_1} \mu_2}{2}.\nonumber
\eea
Similarly, for the $M_P$-even but CP-odd fields, in the basis $(A_H, A_\Phi, A_{\varphi_1}, A_{\varphi_2}, A_{\varphi_3})$, we have the symmetric matrix
\bea\label{eq:MA2}
M^2_{A} = \frac{1}{2}\left(
\begin{array}{ccccc}
 \lambda_2\frac{v_\Phi v_{ \varphi_2} v_{ \varphi_3} }{ v_H} & \star & \star & \star & \star \\
 -\lambda_2 v_{ \varphi_2} v_{ \varphi_3}  & \lambda_2\frac{v_H v_{ \varphi_2} v_{ \varphi_3} }{ v_\Phi} & \star & \star & \star \\
 0 & 0 & \mu_2\frac{v_{ \varphi_2} v_{ \varphi_3} }{ v_{ \varphi_1}} & \star & \star \\
 \lambda_2 v_\Phi v_{ \varphi_3} & - \lambda_2 v_H v_{ \varphi_3} & - \mu_2 v_{ \varphi_3} & \frac{\lambda_2 v_H v_\Phi +\mu_2v_{ \varphi_1}  }{ v_{ \varphi_2}}v_{ \varphi_3} & \star \\
 -\lambda_2 v_\Phi v_{ \varphi_2}  & \lambda_2v_H v_{ \varphi_2}  &  -\mu_2 v_{ \varphi_2} & -\lambda_2 v_H v_\Phi+\mu_2 v_{ \varphi_1} &  \frac{v_H v_\Phi \lambda_2+v_{ \varphi_1}  \mu_2}{v_{ \varphi_3}} v_{ \varphi_2}\\
\end{array}
\right).
\eea
Out of the five eigenvalues of $M_A^2$, only the two below are non-vanishing
\be\label{eq:MA12}
m_{A_{1,2}}^2 = \frac{1}{4 v_H v_\Phi v_{\varphi_1} v_{\varphi_2}v_{\varphi_3}}\left[C_1+C_2\mp\sqrt{(C_1-C_2)^2+4 C_3^2}\right],
\end{equation} 
where
\bea
C_1 &=& \lambda_2 v_{\varphi_1} \left( v_H^2 v_\Phi^2 v_{\varphi_2}^2 + v_H^2 v_\Phi^2v_{\varphi_3}^2 + v_H^2 v_{\varphi_2}^2v_{\varphi_3}^2+v_\Phi^2v_{\varphi_2}^2v_{\varphi_3}^2\right),\\
C_2 &=& \mu_2 v_H v_\Phi \left(v_{\varphi_1}^2v_{\varphi_2}^2+v_{\varphi_1}^2v_{\varphi_3}^2+v_{\varphi_2}^2v_{\varphi_3}^2 \right)\quad \mbox{and} \quad
C_3^2 = \lambda_2 \mu_2v_{H}^3 v_{\Phi}^3 v_{\varphi_1}^3 (v_{\varphi_2}^2-v_{\varphi_3}^2)^2. \nonumber
\eea
Finally, the exact expression for the physical Nambu-Goldstone, $G$, is
\bea\label{eq:G}
G &=& N_G^{-1/2}\left\{ v_H v_\Phi (10 v_{\varphi_1}^2 + 4 v_{\varphi_2}^2 + v_{\varphi_3}^2) (v_\Phi A_H -v_H A_\Phi) + v_{\varphi_1} \left[6 v_H^2 v_\Phi^2 + (v_H^2 + v_\Phi^2)(4 v_{\varphi_2}^2-v_{\varphi_3}^2)\right] A_{\varphi_1} 
 \right.\\
 &&\left.+ v_{\varphi_2} \left[3 v_H^2 v_\Phi^2 - (v_H^2 + v_\Phi^2)(5 v_{\varphi_1}^2+v_{\varphi_3}^2)\right] A_{\varphi_2}
+ v_{\varphi_3} \left[3 v_H^2 v_\Phi^2 + (v_H^2 + v_\Phi^2)(5 v_{\varphi_1}^2+4v_{\varphi_2}^2)\right] A_{\varphi_3}\right\},\nonumber\\
\mbox{with}&&\nonumber\\
\quad N_G &=& \left[9 v_H^2 v_\Phi^2 + (v_H^2 + v_\Phi^2)(25 v_{\varphi_1}^2 + 16v_{\varphi_2}^2+v_{\varphi_3}^2)\right]\left[v_H^2 v_\Phi^2(4 v_{\varphi_1}^2 + v_{\varphi_2}^2+v_{\varphi_3}^2) +  (v_H^2 + v_\Phi^2)(v_{\varphi_1}^2 v_{\varphi_2}^2 + v_{\varphi_1}^2 v_{\varphi_3}^2 + v_{\varphi_2}^2v_{\varphi_3}^2)\right]\nonumber.
\eea

\section{The accidental $U(1)$}\label{app:AccidentalU1}
From the Lagrangian interactions, we can derive relations amongst generic Abelian charges ($Q$) of the different fields in the model. These relations can be expressed in terms of four independent charges. For instance, taking  $B_Q = (Q_{Q_i}, Q_H, Q_{\varphi_1},Q_{\varphi_2}$) as a basis of independent charges, we find the following relations
\begin{eqnarray}\label{eq:Xrels}
Q_{\varphi_3} &=& Q_{\varphi_1} - Q_{\varphi_2},\quad
Q_{\sigma} = \frac{Q_{\varphi_1} + Q_{\varphi_2}}{2},\quad
Q_{\Phi} = Q_H - Q_{\varphi_1} + 2 Q_{\varphi_2}, \\
Q_{\eta} &=& Q_H + \frac{3}{2}\left( Q_{\varphi_2} - Q_{\varphi_1}\right),\quad
Q_{f_a} = -\frac{Q_{\varphi_2}}{2} ,\quad
Q_{N} = \frac{Q_{\varphi_1}}{2} ,\nonumber\\
Q_{L_i} &=& - Q_H +\frac{3}{2} Q_{\varphi_1}-2 Q_{\varphi_2},\quad
Q_{e_i} = -2 Q_H +\frac{3}{2} Q_{\varphi_1}-2 Q_{\varphi_2},\nonumber\\
Q_{d_i} &=&  Q_{Q_i} - Q_H, \quad
Q_{u_i} =  Q_{Q_i} + Q_H. \nonumber
\end{eqnarray}
The four independent charges are associated with four independent global $U(1)$ symmetries, three of which were imposed: $U(1)_Y$, $U(1)_B$ and $U(1)_L$. Thus, by substituting $B_Q = (1/6,1/2,0,0)$, $B_Q = (1/3,0,0,0)$, $B_Q = (0,0,-10,-8)$ into Eq. (\ref{eq:Xrels}) we obtain, respectively, the hypercharge, baryon number and lepton number of all scalars and fermions.
On the other hand, the fourth symmetry, $U(1)_{X}$, was not imposed and appears accidentally in the model. The presence of a massless pseudoscalar in the spectrum tells us about the existence of such a symmetry and that it is broken spontaneously. Note that, as an accidental symmetry, the X-transformation properties of the fields have not been previously defined. Nevertheless, we can obtain such charges {\it a posteriori} by making use of the Goldstone profile given in Eq. (\ref{eq:G}) as well as Goldstone's theorem.

According to Goldstone's theorem, the spontaneous breaking of a continuous global symmetry leads to a massless field, the Goldstone boson, which can be identified through the associated Noether's current. In the case of a global Abelian symmetry, such as the accidental $U(1)_X$, once it is broken by the vevs of a set of scalars $\phi_j = (v_j + S_j + i A_j)/\sqrt{2}$, the corresponding Goldstone boson can be written as
\begin{equation}
G_X = \frac{1}{ \sqrt{\sum_j X_j^2 v_j^2}}\sum_j X_j v_j A_j,
\end{equation}
where $X_j$ represent the charges of $\phi_j$ under $U(1)_X$. Comparing this expression with Eq. (\ref{eq:G}), we can extract the $U(1)_X$ charges of the scalar fields in terms of one of the charges, say $X_H$, and the scalar vevs,
\begin{eqnarray}\label{eq:Xcharges}
\frac{X_{\varphi_1}}{X_H} &=& \frac{6 v_H^2 v_\Phi^2 + (v_H^2 + v_\Phi^2)(4 v_{\varphi_2}^2-v_{\varphi_3}^2)}{v_\Phi^2 (10 v_{\varphi_1}^2 + 4 v_{\varphi_2}^2 + v_{\varphi_3}^2)}
 ,\quad
 \frac{X_{\varphi_2}}{X_H} = \frac{ 3 v_H^2 v_\Phi^2 - (v_H^2 + v_\Phi^2)(5 v_{\varphi_1}^2+v_{\varphi_3}^2)}{v_\Phi^2 (10 v_{\varphi_1}^2 + 4 v_{\varphi_2}^2 + v_{\varphi_3}^2)},\\
 \frac{X_{\varphi_3}}{X_H} &=& \frac{3 v_H^2 v_\Phi^2 + (v_H^2 + v_\Phi^2)(5 v_{\varphi_1}^2+4v_{\varphi_2}^2)}{v_\Phi^2 (10 v_{\varphi_1}^2 + 4 v_{\varphi_2}^2 + v_{\varphi_3}^2)}, \quad
 \frac{X_{\Phi}}{X_H} = - \frac{v_H^2}{v_\Phi^2}.\nonumber
 \end{eqnarray}
The charges of the other fields can be obtained by substituting Eq. (\ref{eq:Xcharges}) with Eq. (\ref{eq:Xrels}).

\bibliographystyle{utphys}
\bibliography{bibliography}
\end{document}